\documentclass[]{aa}
\usepackage{graphicx,natbib}
\usepackage{amssymb}
\DeclareGraphicsExtensions{.eps,.ps}

\def\ms{\mbox{$M_\odot$}}

\begin{document}
\title{Towards a census of the Galactic anticentre star clusters: \\ colour-magnitude diagram and structural analyses \\ of a sample of 50 objects}

\author{D. Camargo\inst{1} \and C. Bonatto\inst{1} \and E. Bica\inst{1}}

\institute{Universidade Federal do Rio Grande do Sul, Departamento de Astronomia, CP\,15051, RS, Porto Alegre 91501-970, Brazil\\
\email{denilso.camargo@ufrgs.br, charles@if.ufrgs.br, bica@if.ufrgs.br}
\mail{denilso.camargo@ufrgs.br}}

\date{Received --; accepted --}

\abstract
{}
{In this work we investigate the nature of 50 overdensities from the catalogue of Froebrich, Scholz, and Raftery (FSR) projected towards the Galactic anticentre, in the sector $160^\circ\,\leq\,\ell\,\leq 200^\circ$. The sample contains candidates with $|b|\,\leq\,20^\circ$  classified by FSR as probable open cluster (OC) and labelled with quality flags 2 and 3. Our main purpose is to determine the nature of these OC candidates and the fraction of these objects that are unknown OCs, as well as to derive astrophysical parameters (\textit{age, reddening, distance, core and cluster radii}) for the clusters and to investigate the relationship among parameters.}
{The analysis is based on 2MASS $J$, $(J-H)$, and $(J-K_s)$ colour-magnitude diagrams (CMDs), and stellar radial density profiles (RDPs) built with decontamination tools. 
The tools are a field star decontamination algorithm, used to uncover the cluster's intrinsic CMD morphology, and colour-magnitude filters to isolate stars with a high probability of being cluster members.}
{Out of the 50 objects, 16 ($32\%$) are star clusters. We show that 9 ($18\%$) overdensities are new OCs (FSR 735, FSR 807, FSR 812, FSR 826, FSR 852, FSR 904, FSR 941, FSR 953, and FSR 955) and 7 ($14\%$) are previously studied or catalogued OCs (KKC1, FSR 795, Cz 22, FSR 828, FSR 856, Cz 24, and NGC 2234). These are OCs with ages in the range $5$ Myr to $1$ Gyr, at distances from the Sun $1.28\lesssim{d}_{\odot}(kpc)\lesssim5.78$ and Galactocentric distances $8.5\lesssim{R}_{GC}(kpc)\lesssim12.9$. We also derive parameters for the previously analysed OCs Cz 22 and NGC2234. Five ($10\%$) candidates are classified as uncertain cases, and the remaining objects are probable field fluctuations.}
{}


\titlerunning{Galactic anticentre star clusters}

\authorrunning{Camargo, Bonatto \& Bica}

\maketitle

%
%________________________________________________________________________

\section{Introduction}
\label{sec:1}
%
%________________________________________________________________________

It is currently accepted that star formation occurs mainly in clustered environments, such as clusters and associations, rather than in isolation. However, only a very small fraction of old stars are found in bound clusters \citep{Lamers05, Lamers06b}. On the other hand, \citet{deWit05} estimate that nearly $95\%$ of the Galactic O star population is located in clusters or OB associations, or can be kinematically linked with them. In this sense, these structures can be thought of as the fundamental units of star formation in the Galaxy \citep{Lada07}.

The spatial and age distribution of clusters has also played a vital role in our understanding of the Galactic structure. In this context, young open clusters (OCs) are important tracers of recent star formation in galaxies and of the spiral structure in galactic disks \citep{Lada03}. On the other hand, old OCs are excellent probes of early disk evolution, and they provide tracers of the structure, kinematics, and chemistry of the Galactic disk \citep{Friel95}.

Primordial conditions during cluster formation and the location of the parental molecular cloud in the Galaxy play an important role in the fate of a cluster \citep{Schilbach06}. However, the stellar content of a cluster evolves with time, and internal and external interactions affect the properties of individual cluster members (e.g. orbit and spatial location), as well as of the whole cluster as a system (e.g. mass and structure). A thorough review of young clusters, 
focussing particularly on the evolution of the massive ones, can be found in \citet{PZ10}.

\begin{figure*}
   \centering
   \includegraphics[scale=0.55,viewport=0 0 470 460,clip]{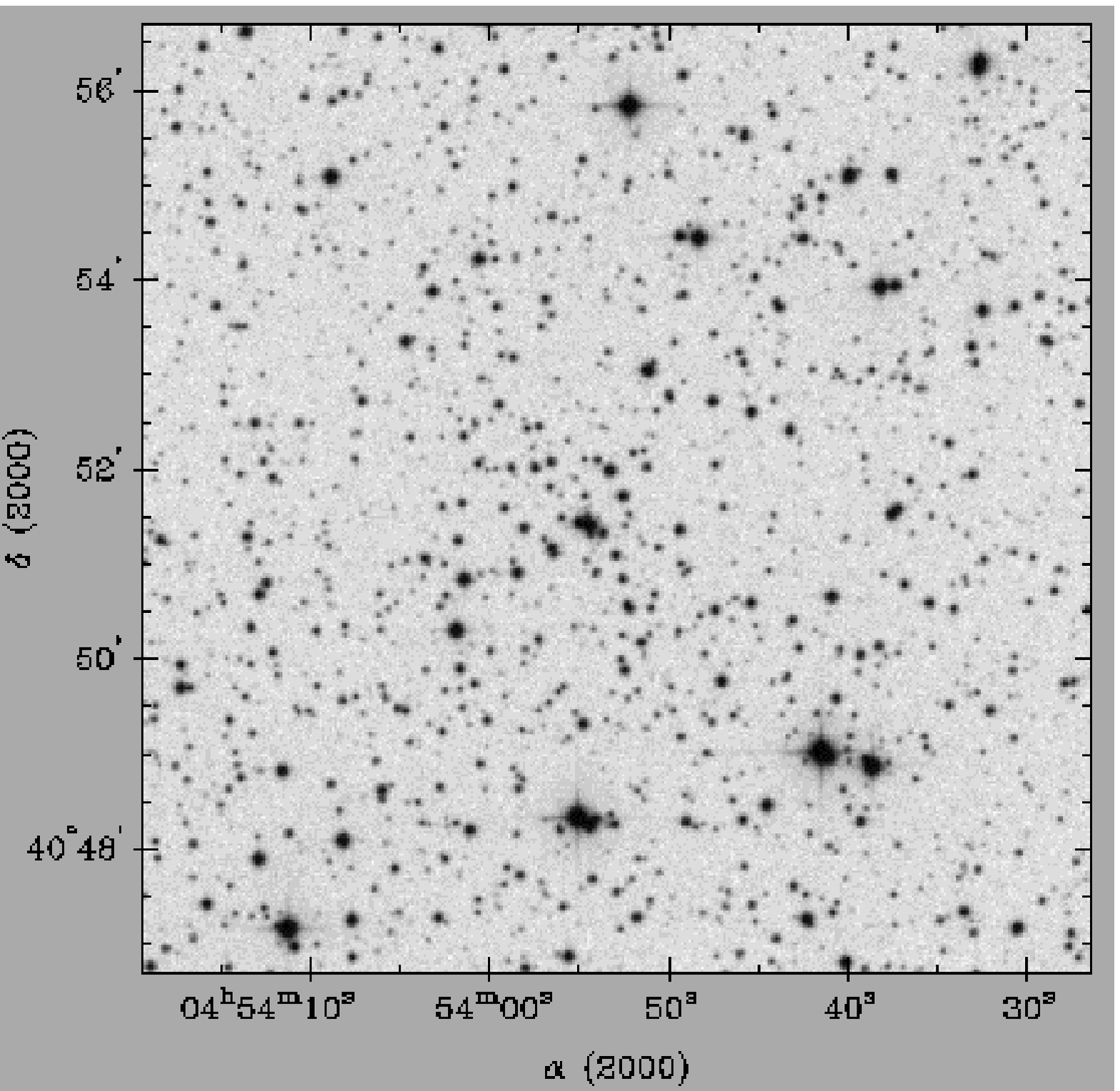}
   \includegraphics[scale=0.55,viewport=0 0 470 460,clip]{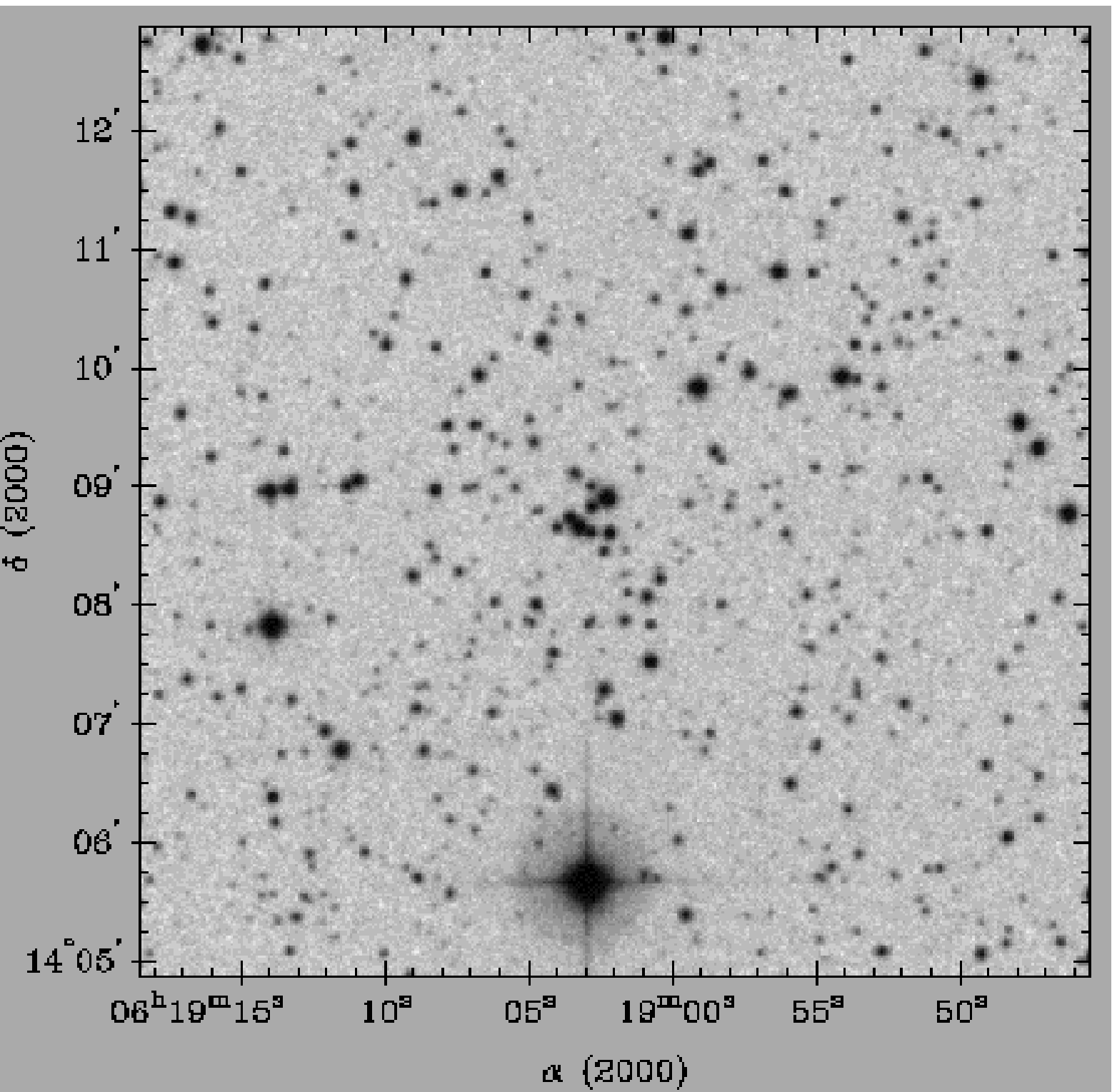}
   \caption[]{Left panel: $10'\times10'$ XDSS R image of FSR $735$. Right panel: $10'\times10'$ XDSS R image of FSR $953$. Images centred on the optimised coordinates.}
   \label{fig:1}
\end{figure*}

\begin{figure*}
\begin{minipage}[b]{0.50\linewidth}
\includegraphics[width=\textwidth]{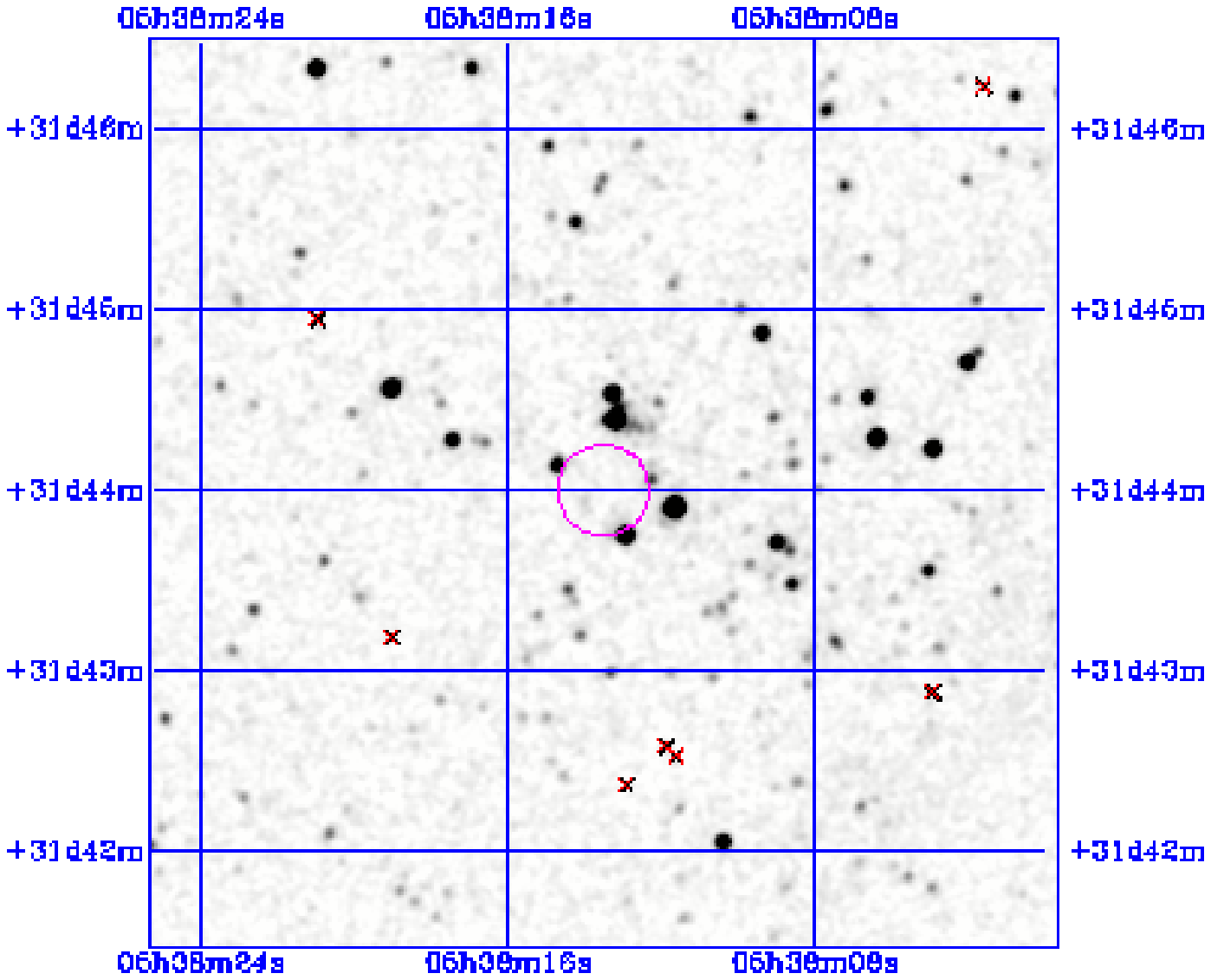}
\end{minipage}\hfill
\begin{minipage}[b]{0.50\linewidth}
\includegraphics[width=\textwidth]{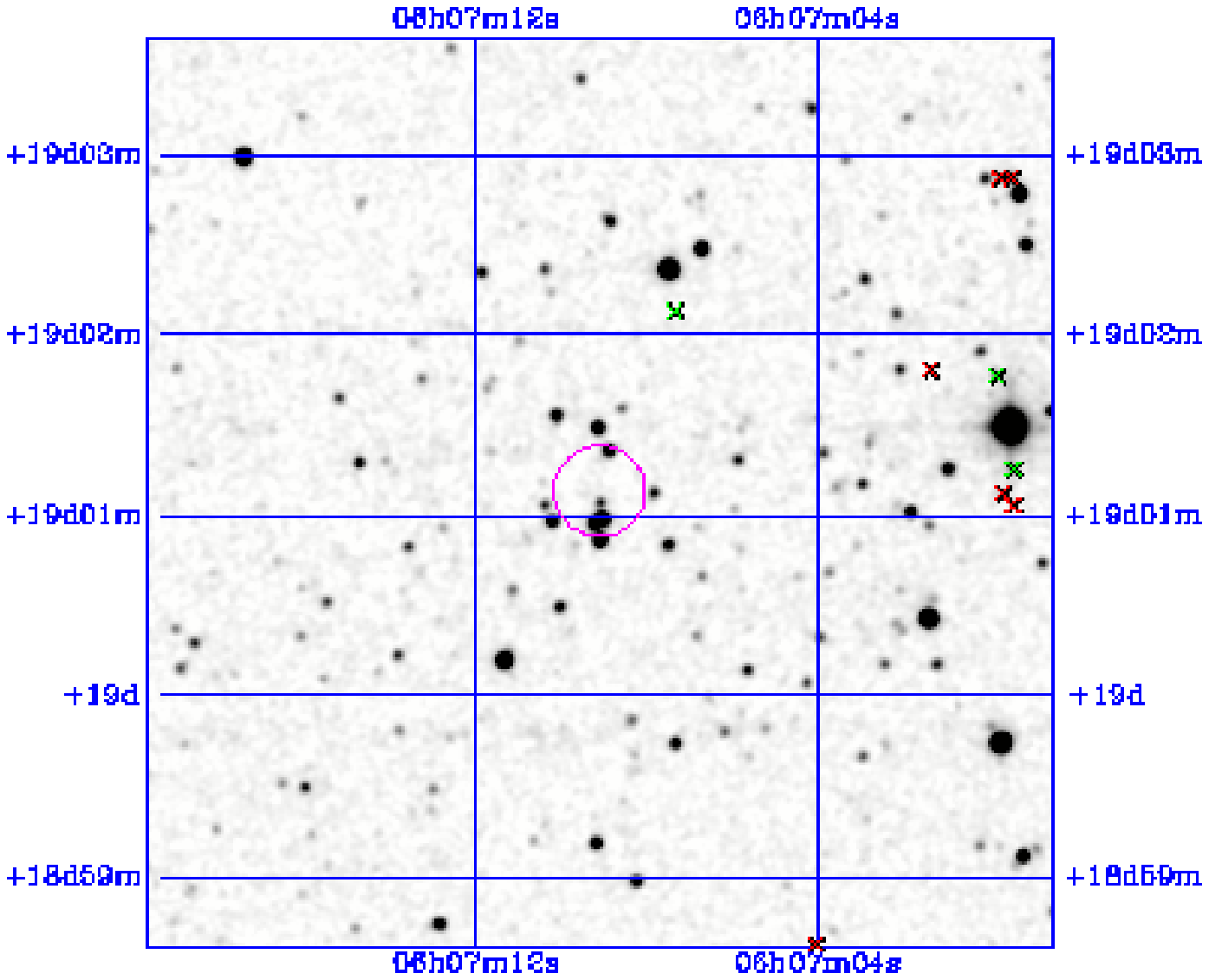}
\end{minipage}\hfill
\caption[]{Left panel: 2MASS K image $5'\times5'$ of FSR 812. Right panel: 2MASS K image $5'\times5'$ of FSR 904. Images centred on the optimised coordinates. The small circle indicates the cluster central region.}
\label{fig:2}
\end{figure*}

The age distribution of star clusters in the Galactic disk can only be explained if they disrupt on a timescale of a few times $10^{8}$ yr \citep[][and references therein]{Lamers06b}. \citet{Lada03} estimate that, in the solar neighbourhood, less than $\sim4\%$ of the clusters formed in giant molecular clouds (GMCs) are able to reach ages beyond 100 Myr, and less than $10\%$ survive longer than $\sim10$ Myr (\textit{infant mortality}). \citet{Oort57} noticed that the distribution of OCs in the solar neighbourhood, as a function of age, shows a lack of old OCs and derived statistically that Galactic clusters disrupt on a timescale of $5\times10^8$ yr. This lack of old clusters can be partially explained by the rapid fading of clusters with age due to stellar evolution, which makes it harder to observe them at older ages. However, fading cannot explain the difference between the observed and the expected number of old OCs, implying that a significant fraction must have been destroyed. \citet{Wielen71} derived a mean dissolution time of 200 Myr from the age distribution of clusters with mass range $10^{2}-10^{3}M_{\odot}$ about 1 kpc from the Sun \citep[see also][]{Lamers06b, Gieles07}. 

Star clusters gradually lose mass and, because of the combined effect of several disruption 
mechanisms\footnote{e.g., mass loss by stellar evolution, mass segregation, and tidal 
interactions with the disk, bulge, spiral arms and GMCs. The latter events increase the 
mean internal energy that may lead to the gradual cluster expansion and disruption.}, they generally end completely destroyed or leave remnants \citep[][and references therein]{Pavani07}. As a consequence, only the more massive OCs (which are essentially gravitationally bound) and those located at large Galactic radii (where the probability of encounters with GMCs is lower) can live a few Gyr \citep{Bergond01}. 

Most of the young star clusters dissolve very early because the primordial gas is removed on a timescale shorter than a crossing time, by winds of massive OB stars and supernova explosions \citep{Tutukov78}. This mechanism strongly depends on the effective star formation efficiency, but appears to be independent of the initial cluster mass \citep{Gieles05}. 

Open clusters are born within GMCs and remain embedded in the clouds for about $2-5$ Myrs. During formation and the earliest stages of evolution, they are often completely invisible at optical wavelengths, since they are only detected in the infrared because of heavy obscuration by gas and dust \citep{Lada03}. The recent development of infrared array detectors has provided an important impulse to our knowledge of these objects.

In the present work we investigate the nature of 50 stellar overdensities from the catalogue of \citet{Froebrich07} towards the Galactic anticentre, in the sector ($160^\circ\,\leq\,\ell\,\leq 200^\circ$), classified by them as probable OC and labelled with quality flags 2 and 3 (Table \ref{tab1}).

\begin{table}[!ht]
\centering
{\footnotesize
\caption{General data on the FSR star cluster candidates.}
\label{tab1}
\renewcommand{\tabcolsep}{0.98mm}
\renewcommand{\arraystretch}{1,1}
\begin{tabular}{lrrrrrrrr}
\hline
\hline
Target&$\alpha(2000)$&$\delta(2000)$&$\ell$&$b$&$R_C$&$R_t$&$Q$\\
&(h\,m\,s)&$(^{\circ}\,^{\prime}\,^{\prime\prime})$&$(^{\circ})$&$(^{\circ})$&$(')$&$(')$& \\
($1$)&($2$)&($3$)&($4$)&($5$)&($6$)&($7$)&($8$)\\
\hline
FSR\,714 &04:42:46&41:55:19&162.024& -2.745&1.32&6.66&2\\
FSR\,717 &04:46:06&42:08:03&162.267&-2.138&1.08&53.7&2\\
FSR\,719  &04:44:19&41:48:51&162.295&-2.597&1.32&18.12&2\\
FSR\,721  &04:46:21&41:57:07&162.436&-2.221&2.82&5.64&3\\
FSR\,723 &04:44:57&41:30:45&162.602&-2.703&0.42&4.56&3\\
FSR\,732 &05:02:36&42:32:55&163.867&+0.49&1.26&4.98&2\\
FSR\,733 &04:45:08&39:46:32&163.945&-3.808&1.98&5.88&3\\
FSR\,735 &04:53:57&40:50:03&164.208&-1.84&2.34&7.02&2\\
FSR\,753 &05:02:14&36:47:24&168.385&-3.084&0.96&10.38&2\\
FSR\,775 &05:25:35&34:57:27&172.641&-0.32&3.18&6.42&3\\
FSR\,778 &05:14:20&32:47:43&173.079&-3.474&0.3&1.56&3\\
FSR\,$788^\dagger$ &05:41:30&35:48:39&173.688&+2.865&0.78&37.5&3\\
FSR\,795 &05:47:29&35:25:56&174.647&+3.707&1.08&54.6&2\\
FSR\,807 &05:36:39&31:49:20&176.529&-0.111&2.04&16.56&2\\
FSR\,812 &05:38:11&31:44:03&176.777&+0.115&1.26&3.84&2\\
FSR\,815 &05:37:42&31:19:27&177.069&-0.19&1.44&4.38&2\\
FSR\,821 &05:41:53&29:54:17&178.749&-0.184&4.98&9.96&3\\
FSR\,$825^\dagger$ &05:48:57&30:10:24&179.317&+1.261&0.66&33.36&2\\
FSR\,826 &05:42:52&28:56:29&179.681&-0.509&0.96&3.90&2\\
FSR\,828 &05:52:18&29:54:16&179.919&+1.746&1.38&43.56&2\\
FSR\,852 &05:53:35&25:10:52&184.133&-0.408&1.56&6.30&3\\
FSR\,856 &06:08:55&26:15:59&184.899&+3.13&0.54&11.04&2\\
FSR\,858 &06:09:09&25:40:38&185.441&+2.891&0.9&41.28&2\\
FSR\,$881^\dagger$ &05:55:25&20:52:59&188.056&-2.216&0.6&31.02&3\\
FSR\,883 &06:04:20&22:00:52&188.106&+0.147&1.62&81.42&2\\
FSR\,891 &06:17:29&22:25:38&189.209&3.011&0.72&5.76&2\\
FSR\,900 &06:06:30&19:13:59&190.78&-0.771&0.3&1.50&3\\
FSR\,902 &06:18:19&20:31:42&190.978&+2.288&0.6&31.5&3\\
FSR\,903 &06:07:34&19:06:46&191.007&-0.611&0.66&21.24&2\\
FSR\,904 &06:07:00&19:00:43&191.03&-0.777&1.38&57.84&2\\
FSR\,906 &06:07:44&18:47:26&191.308&-0.733&2.16&6.42&3\\
FSR\,909 &06:15:46&19:00:49&192.026&+1.039&0.24&1.56&3\\
FSR\,921 &06:05:17&16:40:41&192.87&-2.27&1.14&55.5&2\\
FSR\,924 &06:10:26&16:50:13&193.329&-1.116&1.38&5.58&3\\
FSR\,928 &06:04:20&15:24:50&193.861&-3.09&0.84&40.5&3\\
FSR\,931 &06:13:07&15:56:37&194.422&-0.981&0.3&7.14&3\\
FSR\,933 &06:05:37&14:32:07&194.781&-3.247&4.32&8.70&3\\
FSR\,936 &06:28:31&16:55:36&195.293&+2.737&0.66&33.2&3\\
FSR\,937 &06:13:41&14:59:47&195.319&-1.315&0.48&2.94&2\\
FSR\,$938^\dagger$ &06:29:20&16:45:27&195.535&+2.835&1.26&3.8&2\\
FSR\,939 &06:10:57&14:22:57&195.54&-2.191&0.9&46.0&3\\
FSR\,940 &06:14:16&14:48:18&195.554&-1.284&0.7&37.1&2\\
FSR\,941 &06:21:51&15:45:59&195.574&+0.782&3.0&9.0&2\\
FSR\,951 &06:22:19&14:42:44&196.558&+0.387&0.4&22.4&2\\
FSR\,952 &06:10:15&13:00:47&196.661&-2.997&0.8&10.6&3\\
FSR\,953 &06:19:02&14:08:53&196.681&-0.58&0.4&8.8&2\\
FSR\,955 &06:23:52&14:30:26&196.915&+0.622&0.2&12.5&3\\
FSR\,960 &06:09:53&12:20:02&197.215&-3.401&2.3&4.6&3\\
FSR\,962 &06:23:50&14:05:10&197.284&+0.419&0.8&23.8&3\\
FSR\,975 &06:40:00&13:17:58&199.795&+3.54&2.3&4.5&2\\
\hline
\end{tabular}
\begin{list}{Table Notes.}
\item Cols. $2-3$: Central coordinates provided by \citet{Froebrich07}. Cols. $4-5$: Corresponding Galactic coordinates. Cols. $6-7$: Core and tidal radii derived by \citet{Froebrich07} from King fits. Col. $8$: FSR quality flag. $\dagger$: has a previous designation (Table \ref{tab2}).
\end{list}
}
\end{table}

This paper is organised as follows. In Sect. \ref{sec:2} we present the OC candidates. In Sect. \ref{sec:3} we present the 2MASS photometry and discuss the methods and tools employed in the CMD analyses, especially the field star decontamination algorithm. Section \ref{sec:4} is dedicated to the discussion of the methods and tools used for the analysis of cluster structure. In Sect. \ref{sec:5} we present the results of our analyses of the OC candidates, and derive astrophysical parameters (\textit{age, reddening, distance, core and cluster radii}) of the confirmed OCs and two previously studied objects. In Sect.~\ref{Mass} we estimate the cluster mass stored
in stars. In Sect. \ref{sec:6} we discuss the results and investigate the relationship between derived parameters. Finally, in Sect. \ref{sec:7} we present concluding remarks.

\section{The OC candidates}
\label{sec:2}

\citet{Froebrich07} have published a catalogue of 1021 star cluster candidates with $|b|\,\leq\,20^\circ$ and all Galactic longitudes. They identify overdensities in the 2MASS database that are classified according to a quality flag, with 0 and 1 for the most probable star clusters and 2 - 5 for possible star clusters. \citet{Bica08} explored FSR overdensities, with quality flags 0 and 1, in bulge/disk directions at $|\ell|\,\leq\,60^\circ$. The sample consisted of 20 star cluster candidates and resulted in 4 new, 2 previously known OCs, 5 uncertain cases, and 9 probable field fluctuations. \citet{Bonatto08} analysed 28 FSR cluster candidates projected nearly towards the anti-centre ($160^\circ\,\leq\,\ell\,\leq 200^\circ$) and confirm 6 new and 9 previously known OCs, 6 uncertain cases, and 7 probable fluctuations of the stellar field. 

\begin{table}[!ht]
{\footnotesize
\caption{Cross-identification of the open clusters.}
\renewcommand{\tabcolsep}{1.0mm}
\renewcommand{\arraystretch}{1.1}
\begin{tabular}{lrrrrrrr}
\hline
\hline
Desig\#1&Desig\#2&Desig\#3&Desig\#4&Ref.&Par.\\
($1$)&($2$)&($3$)&($4$)&($5$)&($6$)\\
\hline
KKC1&FSR\,788&Sh2-235\,East1&-&$2,4$&-\\
FSR\,795&Koposov\,10 &-&-&$2,3$&$3$\\
Czernik\,22&Basel\,4&FSR\,825 &OCI\,-\,455&$1,2,3$&-\\
FSR\,828 &Koposov\,43&-&-&$2,3$&$3$\\
FSR\,856 &Koposov\,53&-&-&$2,3$&$3$\\
Czernik\,24 &FSR\,881&OCI\,-\,472&-&$1,2,3$&$3$\\
NGC\,2234&FSR\,938&-&-&$1,2$&-\\
\hline
\end{tabular}
\begin{list}{Table Notes.}
\item Col. $(5)$ show references for cross-identification and col. $(6)$ references for parameters determination. The references are: 1 - \citet{Alter70}; 2 - \citet{Froebrich07}; 3 - \citet{Koposov08};  4 - \citet{Kumar06}.
\end{list}
\label{tab2}
}
\end{table}

\begin{table}[!ht]
\centering
{\footnotesize
\caption{Field star decontamination statistics.}
\label{tab3}
\renewcommand{\tabcolsep}{1.1mm}
\renewcommand{\arraystretch}{1.1}
\begin{tabular}{lrrrrrrr}
\hline
\hline
Target&$(R_{max})$&$N_{obs}$&$N_{cl}$&$N_{1\sigma}$&$\sigma_{FS}$&$FS_{unif}$\\
&$(\,^{\prime}\,)$&(stars)&(stars)&&(stars)& \\
($1$)&($2$)&($3$)&($4$)&($5$)&($6$)&($7$)\\
\hline
\multicolumn{7}{c}{\,\,\,\,\,Confirmed OCs}\\
\cline{2-7}
Cz\,22 &$3$&$177\pm13$&$93\pm3$&$6.5$&$11.7$&$0.09$\\
FSR\,735 &$3$&$175\pm13$&$84\pm3$&$6.1$&$3.0$&$0.02$\\
FSR\,807 &$4$&$150\pm12$&$93\pm3$&$7.1$&$4.8$&$0.04$\\
FSR\,812 &$3$&$92\pm10$&$56\pm2$&$5.7$&$2.2$&$0.03$\\
FSR\,826&$5$&$423\pm21$&$98\pm33$&$4.8$&$8.7$&$0.02$\\
FSR\,852 &$4$&$359\pm19$&$122\pm4$&$6.3$&$10.5$&$0.04$\\
FSR\,904 &$5$&$196\pm14$&$104\pm4$&$7.1$&$6.5$&$0.05$\\
FSR\,941 &$3$&$99\pm10$&$58\pm2$&$5.8$&$3.8$&$0.07$\\
FSR\,953 &$5$&$387\pm20$&$92\pm4$&$3.9$&$24.4$&$0.07$\\
FSR\,955 &$3$&$141\pm12$&$57\pm2$&$4.3$&$6.8$&$0.06$\\
NGC\,2234 &$4$&$250\pm16$&$84\pm4$&$5.0$&$18.2$&$0.08$\\
\cline{2-7}
\multicolumn{7}{c}{\,\,\,\,\,Uncertain cases}\\
\cline{2-7}
FSR\,815 &$2$&$45\pm7$&$32\pm1$&$4.7$&$1.8$&$0.06$\\
FSR\,883 &$3$&$196\pm14$&$62\pm4$&$4.1$&$7.3$&$0.05$\\
FSR\,902 &$2$&$97\pm10$&$42\pm2$&$4.3$&$2.9$&$0.05$\\
FSR\,921 &$3$&$90\pm9$&$44\pm2$&$4.5$&$4.4$&$0.08$\\
FSR\,951 &$2$&$105\pm10$&$50\pm2$&$4.8$&$7.9$&$0.13$\\
\cline{2-7}
\multicolumn{7}{c}{\,\,\,\,\,Possible field fluctuations}\\
\cline{2-7}
FSR\,714 &$3$&$126\pm11$&$34\pm3$&$1.9$&$16.5$&$0.14$ \\
FSR\,717 &$2$&$35\pm6$&$24\pm1$&$4.0$&$1.7$&$0.09$ \\
FSR\,719  &$2$&$62\pm8$&$22\pm1$&$2.7$&$6.9$&$0.15$\\
FSR\,721  &$3$&$88\pm9$&$37\pm2$&$3.8$&$8.6$&$0.15$\\
FSR\,723 &$2$&$74\pm9$&$33\pm1$&$3.7$&$2.6$&$0.05$\\
FSR\,732 &$3$&$90\pm9$&$37\pm3$&$3.7$&$19.6$&$0.30$\\
FSR\,733 &$3$&$119\pm11$&$35\pm2$&$3.0$&$3.0$&$0.03$\\
FSR\,753 &$3$&$58\pm8$&$27\pm1$&$3.4$&$3.4$&$0.09$\\
FSR\,775 &$5$&$362\pm19$&$81\pm4$&$2.0$&$42.6$&$0.28$\\
FSR\,778 &$2$&$40\pm9$&$29\pm1$&$3.9$&$2.8$&$0.13$\\
FSR\,821 &$5$&$391\pm20$&$81\pm5$&$3.6$&$44.5$&$0.12$\\
FSR\,858 &$3$&$192\pm14$&$53\pm2$&$1.0$&$12.5$&$0.07$\\
FSR\,891 &$3$&$156\pm12$&$54\pm3$&$3.5$&$4.2$&$0.03$\\
FSR\,900 &$3$&$171\pm13$&$50\pm3$&$3.6$&$9.1$&$0.07$\\
FSR\,903 &$2$&$87\pm9$&$35\pm2$&$3.7$&$2.7$&$0.05$\\
FSR\,906 &$2$&$87\pm9$&$36\pm2$&$3.7$&$6.5$&$0.11$\\
FSR\,909 &$2$&$82\pm9$&$38\pm2$&$3.3$&$7.1$&$0.12$\\
FSR\,924 &$2$&$83\pm9$&$36\pm2$&$3.9$&$3.4$&$0.06$\\
FSR\,928 &$2$&$43\pm7$&$10\pm1$&$0.6$&$1.5$&$0.03$\\
FSR\,931 &$3$&$154\pm12$&$50\pm3$&$3.8$&$9.1$&$0.08$\\
FSR\,933 &$3$&$130\pm11$&$35\pm3$&$2.8$&$6.2$&$0.06$\\
FSR\,936 &$2$&$74\pm9$&$29\pm2$&$3.3$&$2.1$&$0.04$\\
FSR\,937 &$3$&$154\pm12$&$46\pm3$&$3.2$&$5.5$&$0.04$\\
FSR\,939 &$2$&$82\pm9$&$36\pm2$&$4.0$&$2.9$&$0.05$\\
FSR\,940 &$3$&$123\pm11$&$18\pm2$&$0.9$&$4.9$&$0.10$\\
FSR\,952 &$2$&$50\pm7$&$28\pm1$&$3.9$&$1.9$&$0.07$\\
FSR\,960 &$3$&$156\pm12$&$52\pm3$&$4.0$&$6.5$&$0.05$\\
FSR\,962 &$3$&$138\pm12$&$30\pm3$&$1.5$&$18.4$&$0.14$\\
FSR\,975 &$3$&$154\pm12$&$46\pm3$&$3.6$&$8.0$&$0.07$\\
\hline
\end{tabular}
\begin{list}{Table Notes.}
\item The statistics of the full magnitude range covered by the respective CMD. CMDs extracted from $0\lesssim{R}(')\lesssim{R}_{max}$. 

\end{list}
}
\end{table}

The present FSR overdensity sample is listed in Table \ref{tab1}. Some of these objects have previous identifications (Table \ref{tab2}). The \citet{Koposov08} analysis of CMDs for 5 clusters is showed in Table \ref{tab2}. They concluded that FSR 795 (Koposov 10), Cz 22, FSR 828 (Koposov 43), FSR 856 (Koposov 53) and Cz 24 are clusters. The derived parameters for them and the present ones (Tables \ref{tab3} and \ref{tab4} ) agree well. \citet{Kumar06} studied KKC1 (FSR 788) and found that it is a cluster (Sect. \ref{sec:5}).  We derive parameters for the previously catalogued OCs Cz 22 and NGC2234. \citet{Yadav04} derived for Cz 22 a radius of 1.8', metallicity of 0.08, $E(B-V)=0.45\pm0.05$, age of the $200\pm50$ Myr and $d_{\odot}=3.0\pm0.2$ kpc. NGC2234 does not have parameters derived so far.

Some clusters are seen in visible bands (Fig. \ref{fig:1}), while others are essentially infrared objects (Fig. \ref{fig:2}).

\section{2MASS photometry}
\label{sec:3}

2MASS\footnote{The Two Micron All Sky Survey, available at \textit{www..ipac.caltech.edu/2mass/releases/allsky/}} photometry \citep{Skrutskie06} in the $J$, $H$, and $K_s$ bands was extracted in circular regions centred on the coordinates of the FSR objects using VizieR\footnote{http://vizier.u-strasbg.fr/viz-bin/VizieR?-source=II/246.}. Wide extraction areas are essential for producing RDPs (Sect.\ref{sec:4} ) with a high contrast against the background and for a consistent field-star decontamination (Sect.\ref{sec:3.1}). We started by assuming the FSR coordinates to centre the photometry extraction. Next, we computed the RDP (Sect. \ref{sec:4}) to check cluster centring.  In some cases the RDP built with the original FSR coordinates presented a dip at the centre. Then, new central coordinates are searched (after
field-star decontamination - Sect.~\ref{sec:3.1}) to maximise the star-counts in the innermost 
RDP bin (e.g. \citealt{Bonatto09c}).

\subsection{Field-star decontamination}
\label{sec:3.1}

To uncover the intrinsic CMD morphology from the background stars, we applied the field-star decontamination procedure. This algorithm works on a statistical basis  by measuring the relative number densities of probable cluster and field stars in cubic CMD cells that have axes along the $J$, $(J-H)$ and $(J-K_{s})$ magnitude and colours. These are the colours that provide the maximum distinction among CMD sequences for star clusters of different ages \citep[e.g.][]{Bonatto04}. 

The algorithm (i) divides the full range of magnitude and colours of a given CMD into a 3D grid, (ii) computes the expected number-density of field stars in each cell based on the number of comparison field stars (within  $1\sigma$ Poisson fluctuation) with magnitude and colours compatible with those of the cell, and (iii) subtracts from each cell a number of stars that correspond to the number-density of field stars measured within the same cell in the comparison field. This method is sensitive to local variations in field star contamination with magnitude and colours. Cell dimensions are $\Delta{J}=1.0$, and  $\Delta(J-H)={\Delta(J-K_{s})}=0.2$, which are adequate to allow sufficient star-count statistics in individual cells and preserve the morphology of the CMD evolutionary sequences. The dimensions of the colour/magnitude cells can be changed so that the total number of stars subtracted throughout the whole cluster area matches the expected one, within the $1\sigma$ Poisson fluctuation. We provide here only a brief description of the decontamination procedure. For details see \citet{Bonatto07a} and \citet{Bica08}.

The decontamination algorithm provides the parameters $N_{obs}$, $N_{cl}$, $N_{1\sigma}$, $\sigma_{FS}$, and $FS_{unif}$ \citep{Bica08}, where $N_{obs}$ is the number of observed stars within the spatial region sampled in the CMD, $N_{cl}$ represents the number of probable members after decontamination, and the parameter $N_{1\sigma}$ gives a measure of the statistical significance of the decontaminated number of stars, and  corresponds to the ratio of the number of stars in the decontaminated CMD with respect to the $1\sigma$ Poisson fluctuation measured in the observed CMD. By definition, CMDs of overdensities must have $N_{1\sigma}>1$. It is expected that CMDs of star clusters have integrated $N_{1\sigma}$ that is significantly higher than 1. The $N_{1\sigma}$ values for the present sample are given in col. 5 of Table \ref{tab3}, and $\sigma_{FS}$ corresponds to the ${1\sigma}$ Poisson fluctuation around the mean of the star counts measured in the 8 equal area sectors of the comparison field. Uniform comparison fields present low values of $\sigma_{FS}$. Ideally, star clusters should have $N_{cl}$ higher than $\sim3\sigma_{FS}$. Finally, $FS_{unif}$ measures the star-count uniformity in the comparison field.

\begin{figure}
\resizebox{\hsize}{!}{\includegraphics{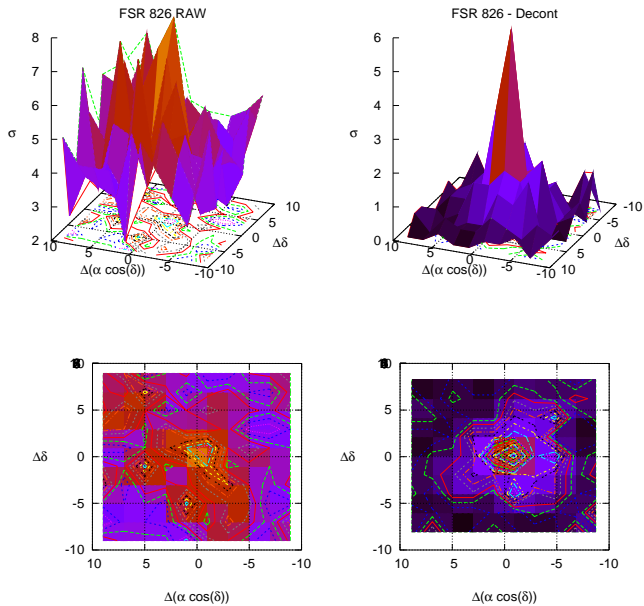}}
\caption[]{Top panels: stellar surface-density $\sigma (stars\,\rm arcmin^{-2}$) of the confirmed OC FSR 826 computed for a mesh size of $3'\times3'$, centred on the coordinates of this object (Tables \ref{tab1} and \ref{tab4}). Bottom: the corresponding isopleth surfaces for observed (raw) photometry and decontaminated photometry.}
\label{fig:3}
\end{figure}

\begin{figure}
\resizebox{\hsize}{!}{\includegraphics{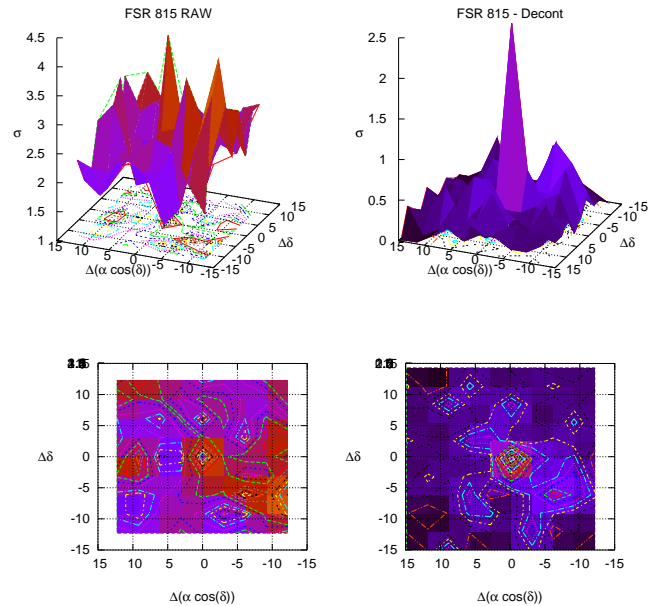}}
\caption[]{Same as Fig. \ref{fig:3} for the uncertain case FSR 815.}
\label{fig:4}
\end{figure}

\begin{figure}
\resizebox{\hsize}{!}{\includegraphics{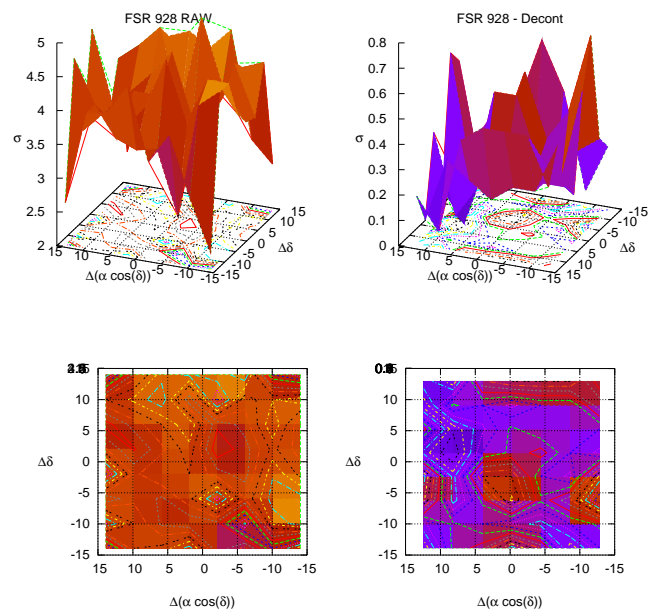}}
\caption[]{Same as Fig. \ref{fig:3} for the possible field fluctuation case FSR 928.}
\label{fig:5}
\end{figure}

\begin{figure}
\resizebox{\hsize}{!}{\includegraphics{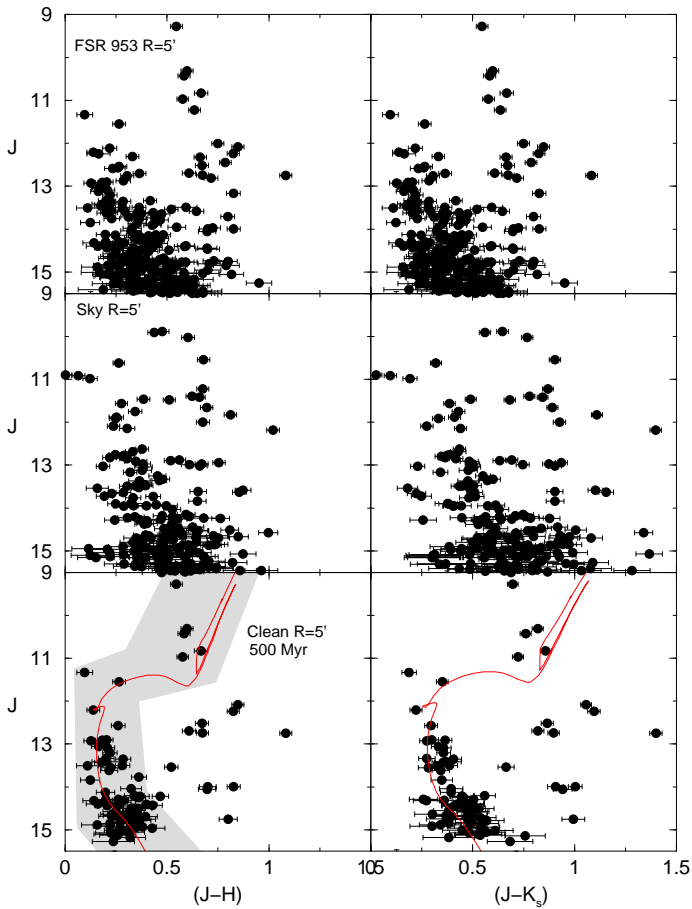}}
\caption[]{2MASS CMDs extracted from the $R=5'$ region of FSR 953. Top panels: observed CMDs $J\times(J-H)$ (left) and $J\times(J-K_s)$ (right). Middle panels: equal area comparison field. Bottom panels: field-star decontaminated CMDs fitted with the 500 Myr Padova isochrone (solid line) for FSR 953. The colour-magnitude filter used to isolate cluster MS/evolved stars is shown as a shaded region}
\label{fig:6}
\end{figure}

Since we usually work with comparison fields larger than the cluster extractions, the correction for the different spatial areas between field and cluster is expected to result in a fractional number of probable field stars ($n^{cell}_{fs}$) in some cells. Before the cell-by-cell subtraction, the fractional numbers are rounded off to the nearest integer, but limited to the number of observed stars in each cell $n^{cell}_{sub}=NI(n^{cell}_{fs})\leq{n^{cell}_{obs}}$, where NI represents rounding off to the nearest integer. The global effect is quantified by means of the difference between the expected number of field stars in each cell ($n^{cell}_{fs}$) and the actual number of subtracted stars ($n^{cell}_{sub}$). Summed over all cells, this quantity provides an estimate of the total subtraction efficiency of the process,

\vspace{0.5cm}
$f_{sub}=100\times\displaystyle\sum_{cell}n^{cell}_{sub}/\sum_{cell}n^{cell}_{fs}\,\,\,\,(\%)$.
\vspace{0.5cm}

Ideally, the best results would be obtained for an efficiency $f_{sub}\approx100\%$. In the present cases, the adopted grid settings produced subtraction efficiencies higher than $90\%$.
In Figs. \ref{fig:3}, \ref{fig:4}, and \ref{fig:5} we show the spatial distribution of the stellar surface density for representative cases. This figure shows the surface density ($\sigma$, in units of stars $arcmin^{-2}$) for a rectangular mesh with cells of dimensions $3'\times3'$.

\subsection{Fundamental parameters}
\label{sec:3.2}

\begin{figure}
\resizebox{\hsize}{!}{\includegraphics{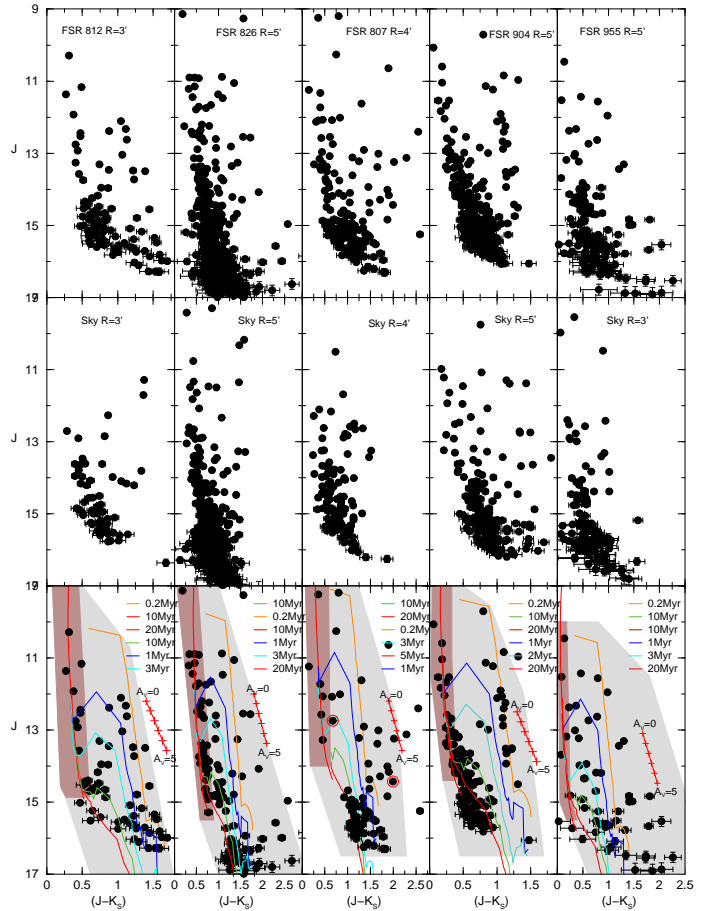}}
\caption[]{2MASS CMDs extracted from the central region of FSR 812, FSR 826, FSR 807, FSR904, and FSR 955. Top panels: observed CMDs $J\times(J-K_s)$. Middle panels: equal area comparison field. Bottom panels: field-star decontaminated CMDs fitted with Padova isochrones (MS) stars and Siess PMS tracks. The colour-magnitude filters used to isolate cluster MS and PMS stars are shown as  shaded regions. We also present the reddening vector for $A_V=0$ to 5. $H\alpha$ emitters of FSR 807 are shown as circles around the stars.}
\label{fig:7}
\end{figure}

The fundamental parameters are derived by means of the constraints provided by the field-decontaminated CMD morphology, especially the combined main sequence (MS) and pre-main sequence (PMS) star distribution, for young OCs (Fig. \ref{fig:7}).
The isochrones from the Padova group with solar metallicity \citep{Girardi02} computed with the 2MASS $J$, $H$, and $K_{s}$ filters are used to represent the MS. Isochrones of \citet{Siess00} are used to characterise the PMS sequences.

\begin{table*}[!ht]
{\footnotesize
\begin{center}
\caption{Derived fundamental parameters.}
\renewcommand{\tabcolsep}{1.4mm}
\renewcommand{\arraystretch}{1.3}
\begin{tabular}{lrrrrrrrrr}
\hline
\hline
Cluster&$\alpha(2000)$&$\delta(2000)$&$E(B-V)$&Age&$d_{\odot}$&$R_{GC}$&$x_{GC}$&$y_{GC}$&$z_{GC}$\\
&(h\,m\,s)&$(^{\circ}\,^{\prime}\,^{\prime\prime})$&(mag)&(Myr)&(kpc)&(kpc)&(kpc)&(kpc)&(kpc)\\
($1$)&($2$)&($3$)&($4$)&($5$)&($6$)&($7$)&($8$)&($9$)&($10$)\\
\hline
\multicolumn{10}{c}{Confirmed OCs}\\
\hline
Cz\,22 &05:48:57&30:10:24&$0.64\pm0.03$&$200\pm50$&$2.6\pm0.1$&$9.8\pm0.2$&$-9.79\pm0.12$&$+0.03\pm0.01$&$-0.06\pm0.01$\\
FSR\,735 &04:53:52.9&40:51:42&$0.51\pm0.03$&$500\pm100$&$2.5\pm0.1$&$9.6\pm0.1$&$-9.56\pm0.10$&$+0.67\pm0.03$&$-0.08\pm0.01$\\
FSR\,807 &05:36:34.2&31:51:20&$1.70\pm0.03$&$5\pm3$&$1.3\pm0.1$&$8.5\pm0.1$&$-8.5\pm0.06$&$+0.08\pm0.01$&$+0.00\pm0.01$\\
FSR\,812 &05:38:13.5&31:44:00&$0.8\pm0.03$&$10\pm5$&$3.3\pm0.2$&$10.5\pm0.2$&$-11.3\pm0.20$&$+0.19\pm0.01$&$+0.01\pm0.01$\\
FSR\,826 &05:42:52.4&28:56:29&$1.09\pm0.03$&$10\pm5$&$2.1\pm0.1$&$9.3\pm0.1$&$-9.27\pm0.01$&$+0.01\pm0.01$&$-0.02\pm0.01$\\
FSR\,852 &05:53:35&25:10:52&$0.32\pm0.03$&$1000\pm200$&$2.2\pm0.1$&$9.4\pm0.1$&$-9.43\pm0.01$&$-0.16\pm0.01$&$-0.02\pm0.01$\\
FSR\,904 &06:07:09.1&19:01:08.2&$0.64\pm0.03$&$20\pm10$&$2.2\pm0.1$&$9.4\pm0.1$&$-9.42\pm0.1$&$-0.43\pm0.02$&$-0.03\pm0.01$\\
FSR\,941 &06:21:47.3&15:44:22.7&$0.80\pm0.03$&$500\pm150$&$5.8\pm0.3$&$12.9\pm0.3$&$-12.8\pm0.3$&$-1.55\pm0.07$&$+0.08\pm0.01$\\
FSR\,953 &\dag\,\,:\dag\,\,:\dag\,\,&\dag\,\,:\dag\,\,:\dag\,\,&$0.48\pm0.03$&$500\pm150$&$2.6\pm0.2$&$9.8\pm0.2$&$-9.72\pm0.12$&$-0.75\pm0.04$&$-0.02\pm0.01$\\ 
FSR\,955 &06:23:56&14:30:26&$0.51\pm0.03$&$10\pm5$&$3.7\pm0.2$&$10.8\pm0.2$&$-10.8\pm0.20$&$-1.09\pm0.05$&$+0.04\pm0.01$\\
NGC\,2234 &06:29:20&16:45:27&$1.25\pm0.03$&$50\pm20$&$4.8\pm0.2$&$11.9\pm0.2$&$-11.88\pm0.2$&$-1.29\pm0.06$&$+0.24\pm0.01$\\
\hline
\multicolumn{10}{c}{Uncertain cases}\\
\hline
FSR\, 815 &05:37:42.8&31:18:19&&&&&&&\\
FSR\, 883 &06:04:26&21:58:54&&&&&&&\\
FSR\, 902 &06:18:20&20:30:25&&&&&&&\\
FSR\, 921 &06:05:13.3&16:40:54&&&&&&&\\
FSR\, 951 &\dag\,\,:\dag\,\,:\dag\,\,&\dag\,\,:\dag\,\,:\dag\,\,&&&&&&&\\
\hline
\end{tabular}
\begin{list}{Table Notes.}
\item Cols. 2 and 3: Optimised central coordinates; ($\dag$) indicates same central coordinates as in \citet{Froebrich07}. Col. 4: reddening in the cluster's central region. Col. 5: age, from 2MASS photometry. Col. 6: distance from the Sun. Col. 7: $R_{GC}$ calculated using $R_{\odot}=7.2$ kpc \citep{Bica06} as the distance of the Sun to the Galactic centre. Cols. 8 - 10: Galactocentric components.
\end{list}
\label{tab4}
\end{center}
}
\end{table*}

\begin{figure}
\resizebox{\hsize}{!}{\includegraphics{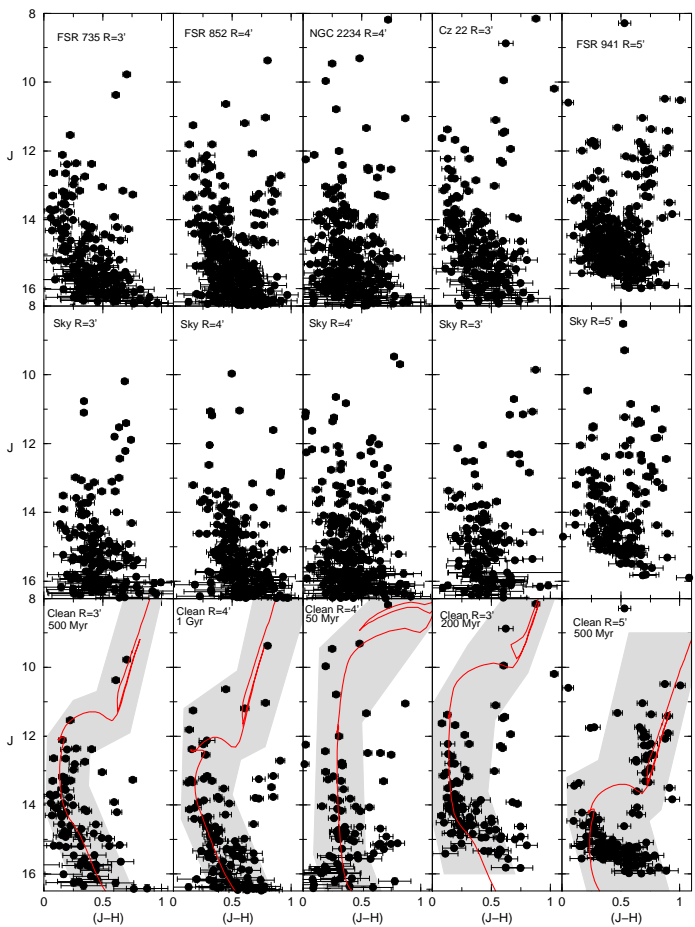}}
\caption[]{2MASS CMDs extracted from the central region of FSR 735, FSR 852, NGC 2234, Cz 22, and FSR 941. Top panels: observed CMDs $J\times(J-H)$. Middle panels: equal area comparison field. Bottom panels: field star decontaminated CMDs fitted Padova isochrone (solid line) for each OC. The colour-magnitude filter used to isolate cluster MS/evolved stars is shown as a shaded region.}
\label{fig:8}
\end{figure}

\begin{figure}
\resizebox{\hsize}{!}{\includegraphics{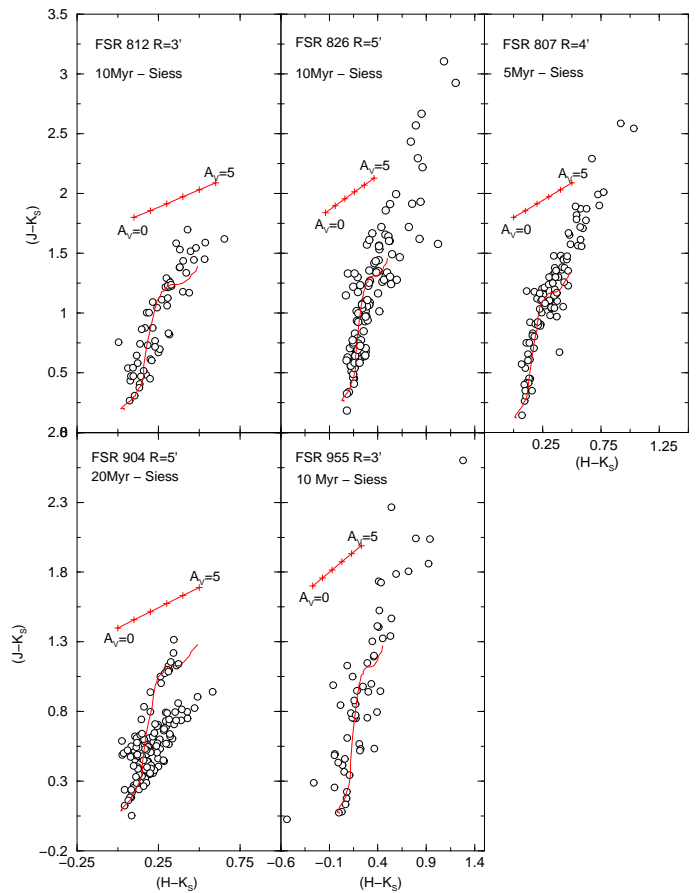}}
\caption[]{Colour-colour diagrams with the decontaminated photometry of the young OCs in our sample,  FSR 812, FSR 826, FSR 807, FSR 904, and FSR 955. \citet{Siess00} isochrones and reddening vectors are used to characterise the PMS distribution.}
\label{fig:9}
\end{figure}

\begin{figure}
\resizebox{\hsize}{!}{\includegraphics{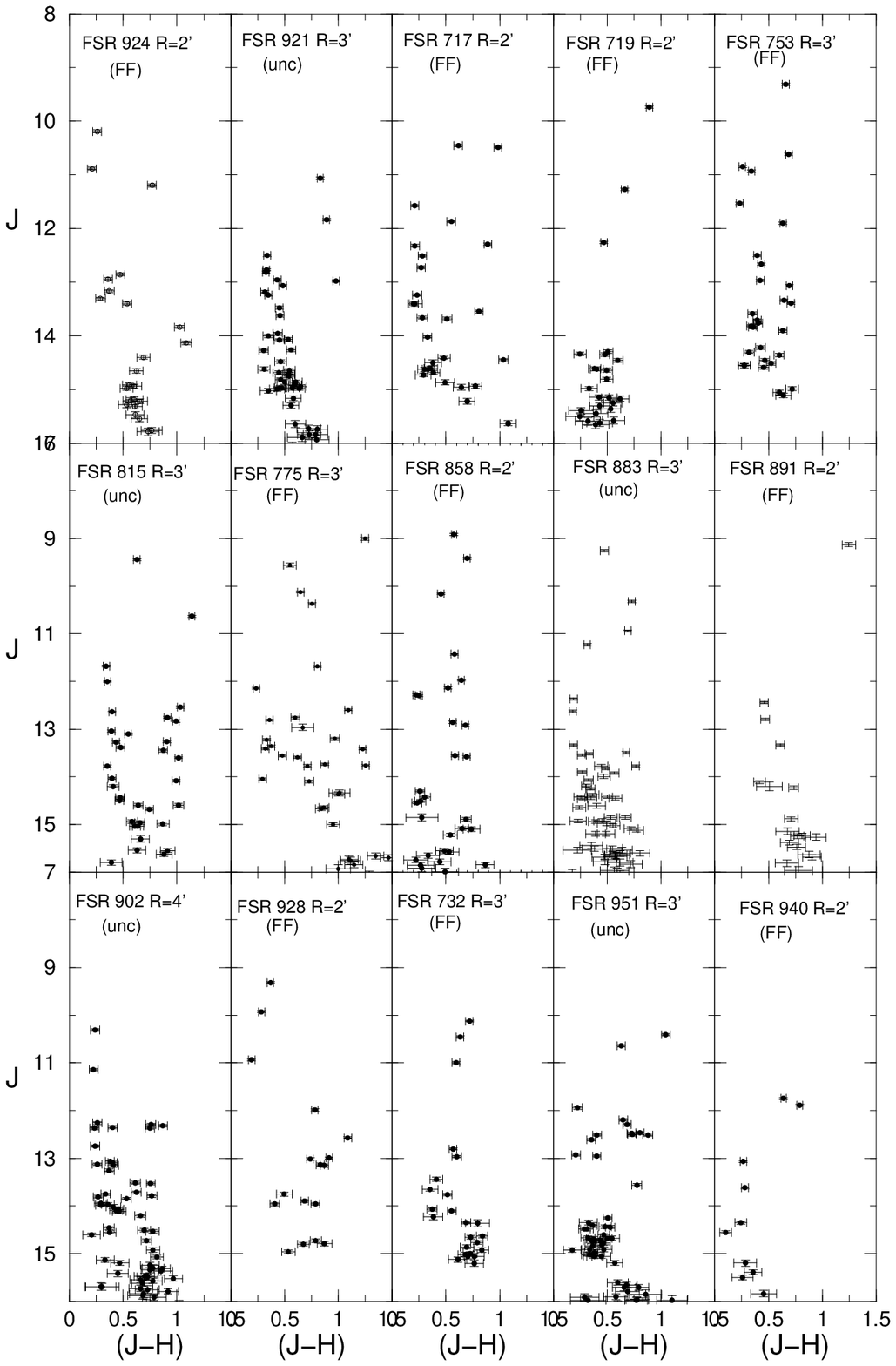}}
\caption[]{Field-star decontaminated $J\times(J-H)$ CMDs of a representative sample of the uncertain cases (unc) and possible field fluctuations (FF).}
\label{fig:10}
\end{figure}

As summarised in \citet{Naylor06}, sophisticated approaches are available for analytical CMD fitting, especially the MS. However, given the poorly-populated MSs, the 2MASS photometric uncertainties for the fainter stars, the important population of PMS stars, and the differential reddening, we directly compared isochrones with the decontaminated CMD morphology. The fits are made by eye, taking the combined MS and PMS stellar distributions as constraints, and allowing for variations due to photometric uncertainties and differential reddening. Specifically, we start with the MS+PMS isochrone set for zero distance modulus and reddening, and next we apply magnitude and colour shifts until a satisfactory solution is reached. The young OCs of this sample present a significant fraction of stars redder than the youngest PMS isochrone. Most of this ($J-K_s$) excess towards the red is probably caused by differential reddening. The best fits are superimposed on decontaminated CMDs (Figs. \ref{fig:6}, \ref{fig:7}, and \ref{fig:8}).
The isochrone fit gives the observed distance modulus $(m-M)_{J}$ and reddening $E(J-H)$, which converts to $E(B-V)$ and $A_{V}$ with the relations $A_{J}/{A_{V}}=0.276$, $A_{H}/{A_{V}}=0.176$, $A_{K_{s}}/{A_{V}}=0.118$, $A_{J}=2.76\times{E(J-H)}$,  and $E(J-H)=0.33\times{E(B-V)}$ \citep{Dutra02}, assuming a constant total-to-selective absorption ratio $R_{V}=3.1$. We  adopt the Sun's distance to the Galactic centre $R_{\odot}=7.2\,kpc$ \citep{Bica06} to compute the Galactocentric distance ($R_{GC}$).
The resulting $E(B-V)$, age, $d_{\odot}$ and $R_{GC}$ are given in cols. 4 to 7 of Table \ref{tab4}. 

\subsection{Colour-colour diagrams}
\label{sec:3.3}

Useful information on the nature of very young OCs can be obtained with colour-colour diagrams. Since our very young OCs include PMS stars, we show in Fig. \ref{fig:9} the decontaminated near-IR colour-colour diagram $(J-K_s)\times(H-K_s)$ of the member stars, together with tracks of \citet{Siess00}, set with the reddening values derived above, to characterise the age. If PMS stars are present in the cluster, it is expected that some stars present near-IR excess. As expected from the CMDs of very young candidates (Fig. \ref{fig:7}), a significant number of the stars appear to be very reddened. Most stars have $(H-K_s)$ colours close to the isochrone, within the uncertainties. Besides, most of the very red PMS stars are displaced parallel to the respective reddening vectors. However, few appear to present an abnormal excess in $(J-K_s)$ and $(H-K_s)$, especially FSR 807, FSR 826, and FSR 955, which may come from PMS stars still bearing circumstellar discs. The cluster can be reddened by foreground, circumcluster cloud, and/or dust around the PMS stars. 

\section{Structural parameters}
\label{sec:4}

Structural parameters are derived by means of the stellar radial density profile (RDP), which is the projected number of stars per area around the centre. RDPs are built with stars selected after applying the respective colour magnitude filter (CM filter) to the observed photometry. These tools isolate the most probable cluster sequences by excluding stars with discordant colours \citep[e.g.][and references therein]{Bonatto07a}. However, residual field stars with colours similar to those of the cluster are expected to remain inside the CM filter region. They affect the intrinsic stellar RDP to a degree that depends on the relative densities of field and cluster stars. The contribution of these residual field stars to the RDPs is statistically quantified by means of a comparison field. In practical terms, the use of the CM filters in cluster sequences enhances the contrast of the RDP to the background. The CM filters are shown in Figs. \ref{fig:6}, \ref{fig:7}, and \ref{fig:8} as the shaded region superimposed on the decontaminated CMDs.

To avoid oversampling near the centre and undersampling for large radii, the RDPs are built 
by counting stars in concentric rings of increasing width with distance to the centre. The 
number and width of rings are optimised so that the resulting RDPs have adequate spatial 
resolution with moderate $1\sigma$ Poisson errors. The residual background level of each 
RDP corresponds to the average number of CM-filtered stars measured in the comparison field.

Usually, the RDPs of star clusters can be described by an analytical profile, like the 
empirical, single mass, modified isothermal spheres of \citet{King66} and \citet{Wilson75}, 
and the power law with a core of \citet{EFF87}. These functions are characterised by different 
sets of parameters that are related to the cluster structure. For simplicity and with the 
RDP error bars (Fig.~\ref{fig:11}), we adopt the two-parameter function $\sigma(R) = \sigma_{bg} + 
\sigma_0/(1+(R/R_c)^2)$, where $\sigma_{bg}$ is the residual background density, $\sigma_0$ the central density of stars, and $R_{core}$ the core radius. Applied to star counts,
this function is similar to that used by \cite{King1962} to describe the surface brightness 
profiles in the central parts of globular clusters. We also estimate the cluster radius 
($R_{RDP}$) by measuring the distance from the cluster centre where the RDP and residual 
background are statistically indistinguishable \citep[e.g.][]{Bonatto07a}. The $R_{RDP}$ can 
be taken as an observational truncation radius, whose value depends both on the radial 
distribution of member stars and the field density.

\section{Results}
\label{sec:5}

The overdensities are classified into three groups, according to the photometric and RDP analyses. 

\subsection{Confirmed open clusters}
\label{sec:5.1}

\begin{figure}
\resizebox{\hsize}{!}{\includegraphics{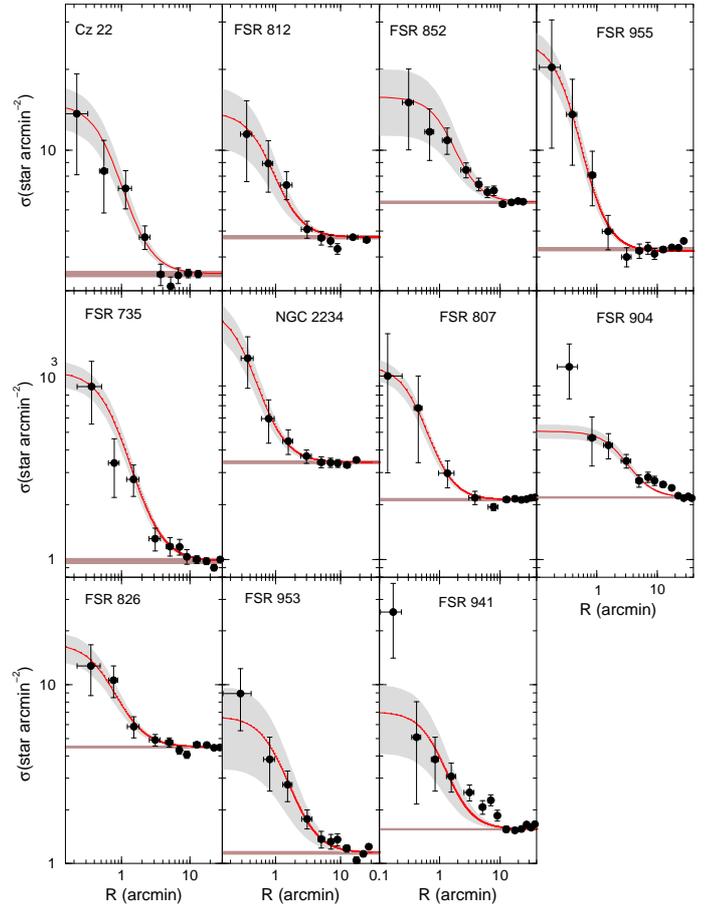}}
\caption[]{Stellar RDPs (filled circles) built with colour-magnitude filtered photometry. Solid line: best-fit King profile. Horizontal shaded region: stellar background level measured in the comparison field. Grey regions: $1\sigma$ King fit uncertainty.}
\label{fig:11}
\end{figure}

This group includes the objects with well-defined decontaminated CMD sequences (Figs. \ref{fig:6}, \ref{fig:7}, and \ref{fig:8}) with relatively high values of the parameter $N_{1\sigma}$, as well as King-like RDPs (Fig. \ref{fig:11}). For young OCs we also built  colour-colour diagrams (Fig. \ref{fig:9}). The astrophysical parameters could be measured for these objects (Tables \ref{tab4} and \ref{tab5}). The previously unknown OCs are FSR 735, FSR 807, FSR 812, FSR 826, FSR 852, FSR 904, FSR 941, FSR 953, and FSR 955. We also derived parameters for the previously catalogued OCs Cz 22 and NGC2234. KKC1 was confirmed as an OC, but the parameters of this object will be analysed in a forthcoming paper.

In Fig. \ref{fig:6} we present the $J\times(J-H)$ and $J\times(J-K_s)$ CMDs extracted from a region $R=5'$ centred on the optimised coordinates of the confirmed OC FSR 953 (top-panel). In the middle panels we show the comparison field corresponding to a ring with the same area as the central region. In the bottom panels we show the decontaminated CMDs with the 500 Myr Padova isochrones fitted.
Figure \ref{fig:7} show $J\times(J-K_s)$ CMDs for the confirmed young OCs FSR 812, FSR 826, FSR 807, FSR 904, and FSR 955. These objects present important populations of PMS stars and, therefore, we also use isochrones of \citet{Siess00}. To examine differential reddening, we include reddening vectors computed with the 2MASS ratios for visual absorptions in the range $A_V=0$ to $5$. 
We present in Fig. \ref{fig:8} the $J\times(J-H)$ CMDs for the remaining confirmed OCs FSR 735, FSR 852, NGC2234, Cz 22 and FSR 941.
In Fig. \ref{fig:11}, we present the RDPs of these objects, and in Table \ref{tab5} we show the derived structural parameters. We show in Table \ref{tab6} integrated colours and magnitudes for confirmed OCs.

Both FSR 904 and FSR 941 present a conspicuous excess over the King-like profile in the innermost RDP bin. This cusp has been detected in post-core collapse globular clusters \citep{Trager95} and some Gyr-old OCs, such as NGC3960 \citep{Bonatto06} and LK 10 \citep{Bonatto09a}. It has been attributed to advanced dynamical evolution. With $\sim500$ Myr of age, FSR 941 is probably a core-collapsed OC. However, some very young OCs also present a cusp, probably as a consequence of molecular cloud fragmentation and/or star formation effects. In this context, we can mention NGC2244 \citep{Bonatto09b}, NGC6823 \citep{Bica08b}, Pismis 5 and NGC1931 \citep{Bonatto09c}, and FSR 198 \citep{Camargo09} as examples of young OCs with a central cusp. FSR 904 presents a similar effect.

\begin{figure}
\resizebox{\hsize}{!}{\includegraphics{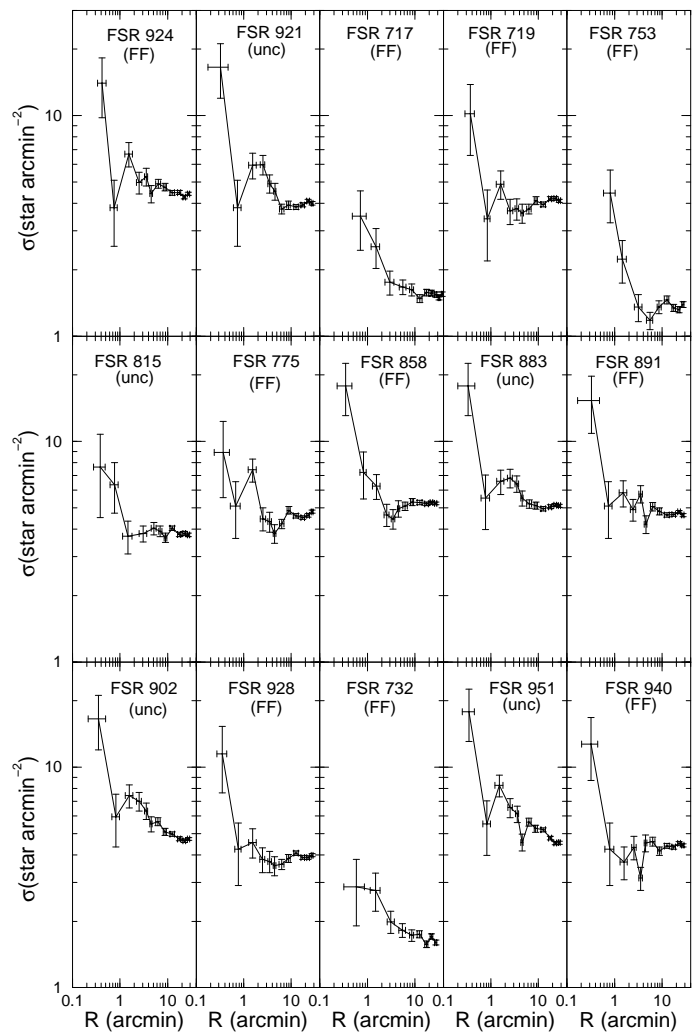}}
\caption[]{RDPs of a representative sample of uncertain cases  and probable field fluctuations.}
\label{fig:12}
\end{figure}

\citet{Witham08} presented the INT/WFC Photometric $H\alpha$ Survey of the Northern Galactic Plane (IPHAS). This catalogue contains positions and photometry for $4853$ sources with $H\alpha$ excess \citep[see also,][]{Drew05}. The $H\alpha$ emission is linked to several events in star clusters as stellar winds, T Tauri stars, Herbig-Haro objects, planetary nebulae and HII region associations. FSR 807 presents 2 $H\alpha$-excess sources within the cluster radius and 3 in the neighbourhood field ($J053635.21+31503.0$ at 1.3' of the cluster centre, $J053631.98+314939.6$ at 1.7', $J053641.46+314627.8$ at 5.1', $J053619.82+314356.2$ at 8.0', and $J053619.58+314353.6$ at 8.06') and FSR 812 presents 1 object ($J053817.88+313934.8$) at $4.5'$ of the cluster centre. All objects are also present in the emission-line star catalogue of \citet{Kohoutek99}. We indicate the $2$ $H\alpha$ emitters in FSR $807$ CMD  Fig. \ref{fig:7}.

\begin{table*}[!ht]
{\footnotesize
\begin{center}
\caption{Structural parameters.}
\renewcommand{\tabcolsep}{0.9mm}
\renewcommand{\arraystretch}{1.3}
\begin{tabular}{lrrrrrrrrrrr}
\hline
\hline
Cluster&$(1')$&$\sigma_{0K}$&$\sigma_{bg}$&$R_{core}$&$R_{RDP}$&$\sigma_{0K}$&$\sigma_{bg}$&$R_{core}$&$R_{RDP}$&${\Delta}R$&CC\\
&($pc$)&($*\,pc^{-2}$)&($*\,pc^{-2}$)&($pc$)&($pc$)&($*\,\arcmin^{-2}$)&($*\,\arcmin^{-2}$)&($\arcmin$)&($\arcmin$)&($\arcmin$)&\\
($1$)&($2$)&($3$)&($4$)&($5$)&($6$)&($7$)&($8$)&($9$)&($10$)&($11$)&($12$)\\
\hline
Cz\,22 &$0.74$&$20.8\pm4.7$&$3.5\pm0.1$&$0.53\pm0.07$&$4.1\pm1.1$&$11.42\pm2.6$&$1.93\pm0.04$&$0.72\pm0.1$&$5.5\pm1.5$&$10-20$&$0.95$\\
FSR\,735 &$0.71$&$19.2\pm3.2$&$2.0\pm0.06$&$0.53\pm0.06$&$6.1\pm0.7$&$9.79\pm1.64$&$0.98\pm0.03$&$0.75\pm0.09$&$8.5\pm1.0$&$10-30$&$0.95$\\
FSR\,807 &$0.37$&$68.2\pm10.8$&$15.5\pm0.3$&$0.15\pm0.01$&$1.48\pm0.4$&$9.41\pm1.5$&$2.14\pm0.04$&$0.41\pm0.04$&$4.0\pm0.5$&$20-40$&$0.97$\\
FSR\,812 &$0.96$&$9.9\pm3.9$&$5.2\pm0.1$&$0.7\pm0.08$&$3.8\pm1.0$&$9.14\pm3.6$&$4.74\pm0.07$&$0.72\pm0.2$&$4.0\pm1.0$&$10-20$&$0.91$\\
FSR\,826 &$0.59$&$35.8\pm8.6$&$12.9\pm0.12$&$0.35\pm0.06$&$3.0\pm0.9$&$12.5\pm3.0$&$4.5\pm0.04$&$0.59\pm0.1$&$5.0\pm1.5$&$20-40$&$0.96$\\
FSR\,852 &$0.64$&$42.5\pm18.7$&$11.4\pm0.09$&$0.6\pm0.2$&$6.4\pm0.6$&$17.42\pm7.69$&$4.68\pm0.04$&$1.01\pm0.27$&$10.0\pm1.0$&$20-30$&$0.90$\\
FSR\,904&$0.65$&$6.8\pm1.0$&$5.2\pm0.05$&$1.71\pm0.27$&$11.0\pm2.0$&$2.88\pm0.43$&$2.19\pm0.02$&$2.63\pm0.41$&$17.0\pm3.0$&$20-40$&$0.95$\\
FSR\,941&$1.68$&$1.9\pm1.15$&$0.6\pm0.02$&$1.44\pm0.70$&$16.0\pm3.4$&$5.50\pm3.25$&$1.584\pm0.01$&$0.86\pm0.42$&$9.5\pm3.0$&$20-40$&$0.80$\\
FSR\,953 &$0.75$&$9.75\pm5.7$&$2.03\pm0.03$&$0.77\pm0.3$&$8.3\pm1.1$&$5.51\pm3.21$&$1.15\pm0.02$&$1.03\pm0.42$&$11.0\pm1.5$&$20-30$&$0.85$\\
FSR\,955&$1.08$&$18.0\pm3.3$&$3.6\pm0.05$&$0.39\pm0.05$&$3.2\pm1.1$&$21.0\pm3.87$&$4.28\pm0.06$&$0.35\pm0.05$&$3.0\pm1.0$&$10-30$&$0.97$\\
NGC\,2234 &$1.40$&$4.0\pm1.8$&$0.9\pm0.01$&$0.78\pm0.2$&$5.6\pm1.4$&$7.94\pm3.5$&$1.73\pm0.03$&$0.56\pm0.17$&$4.0\pm1.0$&$20-30$&$0.91$\\
\hline
\end{tabular}
\begin{list}{Table Notes.}
\item Col. 2: arcmin to parsec scale. To minimise degrees of freedom in RDP fits with the King-like profile (see text), $\sigma_{bg}$ was kept fixed (measured in the respective comparison fields) while $\sigma_{0}$ and $R_{core}$ were allowed to vary. Col. 11: comparison field ring. Col. 12: correlation coefficient.
\end{list}
\label{tab5}
\end{center}
}
\end{table*}

\subsection{Uncertain cases}
\label{sec:5.2}

The objects in this group have, in general, less defined decontaminated CMD sequences than those of the confirmed OCs, which is consistent with the lower level of the integrated $N_{1\sigma}$ parameter. The irregular RDPs make difficult King's law fits.  By ``uncertain cluster'', we mean those objects with a CMD that may suggest a cluster, but not the RDP (or the contrary). We suggest that deeper photometry, proper motions and other methods be employed to explore them in more detail. The uncertain cases are FSR 815, FSR 883, FSR 902, FSR 921, and FSR 951. In Fig. \ref{fig:10} we show the decontaminated $J\times(J-H)$ CMDs of a representative sample of the uncertain cases, and in Fig. \ref{fig:12} their RDPs.

\subsection{Possible field fluctuations}
\label{sec:5.3}

Decontaminated CMDs of this group have $N_{1\sigma}$-values significantly lower, and the RDPs much more irregular, than those of the other two groups (Sect. \ref{sec:3.1}). In Fig.\ref{fig:10} we present the decontaminated $J\times(J-H)$ CMDs of a representative sample of the possible field fluctuations and Fig. \ref{fig:12} shows RDPs for some these overdensities.

\section{Mass estimate}
\label{Mass}

Given the somewhat limited 2MASS photometric depth and the relatively large
distance (Table~\ref{tab5}) of our confirmed OCs, the CMDs in Figs. \ref{fig:6} - \ref{fig:8} do not contain 
the whole mass range expected especially for OCs older than $\sim50$Myr. Thus, we estimate the
stellar mass by means of the mass function (MF), built for the observed MS mass range according 
to \citet{Bica06}. The MS MF is then fitted with the function $\phi(m)\propto{m}^{-(1+\chi)}$. Details of this approach are given in Table \ref{tab7}, where we also show the number and mass of the evolved stars. Clearly, we do not have access to the lower MS. Thus, assuming that the low-mass content is still present, we use Kroupa's (2001) MF\footnote{$\chi=0.3\pm0.5$ for $0.08<m(M_{\odot})<0.5$, $\chi=1.3\pm0.3$ for $0.5<m(M_{\odot})<1.0$, and $\chi=1.3\pm0.7$ for $1.0<m(M_{\odot})$.} to estimate the total stellar mass, down to the H-burning mass limit ($0.08\,M_{\odot}$). The results are given in the last two columns of Table \ref{tab7}. Interestingly, the extrapolation suggests that FSR 941 may be a relatively massive OC.

For the young OCs we built the MS MF in a similar way to the old ones, and count the number of 
PMS stars (Table \ref{tab8}). Interestingly, the MF slopes are, in general, flatter than those 
of the older OCs (Table \ref{tab7}), which may reflect the longer timescale for the evolution
towards the MS of the low-mass PMS stars. Given the differential reddening, it is not possible 
to attribute a precise mass value for each PMS star. Thus, we simply count the number of PMS 
stars and adopt an average mass value for the PMS stars to estimate $n_{PMS}$ and $m_{PMS}$.
Assuming that the mass distribution of the PMS stars also follows Kroupa's (2001) MF, the 
average PMS mass - for masses within the range $0.08\la m(\ms)\la7$ - is $<m_{PMS}>\approx0.6\ms$.
Thus, we simply multiply the number of PMS stars (Table~8) by this value to estimate the PMS 
mass. Finally, we add the latter value to the MS mass to obtain an estimate of the total 
stellar mass. Obviously, similarly to the MS stars, 2MASS cannot detect the very low mass 
PMS stars. Consequently, these values should be taken as lower limits.

\section{Discussion}
\label{sec:6}

After cluster formation, structural parameters change with stellar and dynamical evolution. As a consequence of the rapid expulsion of primordial gas and the new lower gravitational potential, the cluster increases on all scales reaching for virialisation. \citet{Goodwin06} show that this early core radii expansion phase occurs within 10 - 30 Myr and can be explained as an effect of rapid residual gas expulsion. After gas expulsion (a few $10^7$ yr), when some energy equipartition is reached, the core radius of the OC survivors shrink, whereas the outer parts keep increasing in size. Mass loss due to stellar evolution also affects the structural, parameters but this effect is negligible because the most massive stars ($M_{\star}>30M_{\odot}$) hardly contribute to the mass of the cluster \citep{Lamers06b}. In this context, \citet{Portegies99} show that the maximum effect of stellar evolution on core radius expansion is about a factor two, but \citet{Mackey08} show that this effect is more significant if the star cluster is initially mass segregated. They also show that significant core expansion due to stellar evolution occurs on timescales of $\sim100$ Myr.

\begin{table*}
{\footnotesize
\begin{center}
\caption{Integrated colours and magnitudes.}
\label{tab6}
\renewcommand{\tabcolsep}{0.9mm}
\renewcommand{\arraystretch}{1.3}
\begin{tabular}{lrrrrrrrrrrrrrrrrrrrrrr}
\hline
\hline
\multicolumn{8}{c}{Magnitude}&\multicolumn{4}{c}{Colour}&\multicolumn{11}{c}{Half-light\,\&\,Starcount\,Radii}\\
\cline{2-8}
\cline{10-11}
\cline{13-23}
\multicolumn{1}{c}{}&\multicolumn{3}{c}{Apparent}&\multicolumn{1}{c}{}&\multicolumn{3}{c}{Absolute}&\multicolumn{1}{c}{}&\multicolumn{3}{c}{Reddening Corrected}&\multicolumn{1}{c}{}&\multicolumn{3}{c}{$R_{hl}$ (pc)}&\multicolumn{1}{c}{}&\multicolumn{3}{c}{$R_{hl}$(')}&\multicolumn{1}{c}{$R_{sc}$(pc)}&\multicolumn{1}{c}{}&\multicolumn{1}{c}{$R_{sc}(')$}\\
\cline{2-4}
\cline{6-8}
\cline{10-11}
\cline{13-15}
\cline{17-19}
\cline{21-21}
\cline{23-23}
Cluster&$J$&$H$&$K_s$&&$J$&$H$&$K_s$&&$(J-H)$&$(J-K_s)$&&$J$&$H$&$K_s$&&$J$&$H$&$K_s$&&&&\\
(1)&(2)&(3)&(4)&&(5)&(6)&(7)&&(8)&(9)&&(10)&(11)&(12)&&(13)&(14)&(15)&&(16)&&(17)\\
\hline
Cz\,22&7.5&6.9&6.7&&-5.1&-5.5&-5.6&&$0.46\pm0.29$&$0.56\pm0.19$&&0.8&0.8&0.8&&1.1&1.1&1.0&&$1.6\pm0.2$&&$2.2\pm0.3$\\
FSR\,735&8.0&7.4&7.3&&-5.3&-5.4&-5.3&&$0.17\pm0.03$&$0.01\pm0.03$&&2.4&2.4&2.4&&3.4&3.4&3.4&&$1.1\pm0.2$&&$1.5\pm0.3$\\
FSR\,807&8.7&8.5&8.4&&-3.1&-3.1&-3.1&&$0.34\pm0.04$&$0.50\pm0.03$&&0.7&0.7&0.7&&1.4&1.4&1.4&&$0.4\pm0.2$&&$0.8\pm0.5$\\
FSR\,812&9.1&8.8&8.7&&-4.2&-4.2&-4.2&&$0.02\pm0.02$&$0.00\pm0.02$&&1.3&1.3&1.3&&1.4&1.4&1.4&&$1.3\pm0.7$&&$1.4\pm0.7$\\
FSR\,826&9.5&9.0&8.8&&-3.0&-3.1&-3.2&&$0.15\pm0.02$&$0.21\pm0.02$&&1.6&1.6&1.6&&2.7&2.7&2.7&&$1.4\pm0.5$&&$2.4\pm0.9$\\
FSR\,852&8.1&7.6&7.4&&-3.9&-4.3&-4.5&&$0.43\pm0.04$&$0.61\pm0.04$&&5.0&5.0&5.0&&7.8&7.8&7.8&&$2.8\pm0.8$&&$4.4\pm1.2$\\
FSR\,904&7.8&7.7&7.4&&-4.5&-4.4&-4.6&&$0.01\pm0.02$&$0.12\pm0.02$&&2.4&2.4&2.4&&3.7&3.7&3.7&&$5.6\pm0.6$&&$8.6\pm0.9$\\
FSR\,941&$10.5$&$10.2$&$10.0$&&$-4.0$&$-4.1$&$-4.2$&&$0.05\pm0.13$&$0.13\pm0.10$&&1.3&1.2&1.2&&0.8&0.7&0.7&&$9.7\pm0.8$&&$5.8\pm0.5$\\
FSR\,953&8.6&8.2&8.1&&-3.9&-4.1&-4.1&&$0.35\pm0.03$&$0.46\pm0.03$&&2.3&2.3&2.3&&3.0&3.0&3.0&&$4.0\pm1.3$&&$5.2\pm1.8$\\
FSR\,955&12.0&11.8&10.8&&-1.3&-1.3&-2.2&&$0.03\pm0.24$&$0.89\pm0.17$&&0.8&0.8&1.4&&0.8&0.7&1.3&&$0.7\pm0.4$&&$0.6\pm0.4$\\
NGC\,2234&7.2&6.7&6.5&&-5.3&-5.6&-5.7&&$0.36\pm0.11$&$0.44\pm0.28$&&1.6&1.6&1.6&&2.1&2.1&2.1&&$1.9\pm0.1$&&$2.5\pm0.1$\\
\hline
\end{tabular}
\begin{list}{Table Notes.}
\item Col. 2-4: apparent magnitude. Cols. 5-7: absolute magnitude. Cols. 8-9: Reddening-corrected $(J-H)$ and $(J-K_s)$  colours. Cols. 10-15: half-light radii in pc and arcmin. Cols. 16-17: star-count radii in pc and arcmin.
\end{list}
\end{center}
}
\end{table*}

\begin{table*}
\caption[]{Stellar mass estimate for clusters older than 50\,Myr}
\label{tab7}
%\tiny
\renewcommand{\tabcolsep}{2.6mm}
\renewcommand{\arraystretch}{1.25}
\begin{tabular}{ccccccccccc}
\hline\hline
&&\multicolumn{6}{c}{Observed in CMD}&&\multicolumn{2}{c}{Extrapolated}\\
\cline{3-8}\cline{10-11}
Cluster&&$\Delta\,m_{MS}$&$\chi$&$N_{MS}$&$M_{MS}$   &$N_{evol}$&$M_{evol}$ &&$N$&$M$\\
     && ($M_\odot$)    &      &(stars) &($M_\odot$)&(stars)   &($M_\odot$)&&(stars) &($M_\odot$)     \\
(1)&&(2)&(3)&(4)&(5)&(6)&(7)&&(8)&(9)\\
\hline
FSR\,735&&0.55-2.90&$1.21\pm0.11$&$111\pm14$&$164\pm20$&$5\pm1$&$15\pm3$&&$1200\pm889$&$467\pm166$\\
FSR\,852&&0.65-2.10&$2.59\pm0.46$&$132\pm19$&$152\pm22$&$4\pm6$&$8\pm12$&&$2710\pm2190$&$836\pm409$\\
FSR\,953&&0.85-2.50&$2.41\pm0.29$&$108\pm14$&$178\pm21$&$8\pm3$&$16\pm6$&&$5020\pm3850$&$1670\pm719$\\
Cz\,22\,\,&&1.30-3.90&$0.87\pm0.20$&$60\pm10$&$135\pm21$&$5\pm3$&$15\pm11$&&$1790\pm1290$&$777\pm239$\\
FSR\,941&&1.70-2.90&$0.67\pm0.12$&$147\pm13$&$350\pm31$&$40\pm15$&$106\pm39$&&$13800\pm10200$&$5480\pm1910$\\
NGC\,2234&&0.95-6.75&$0.97\pm0.10$&$139\pm8$&$431\pm28$&$4\pm2$&$32\pm18$&&$2640\pm1860$&$1360\pm344$\\
\hline
\end{tabular}
\begin{list}{Table Notes.}
\item Col. 2: MS mass range. Col. 3: MS mass function slope $\chi$, derived from the fit of $\phi(m)\propto{m}^{-(1+\chi)}$. Cols. 4-7: stellar content of the MS and evolved stars. Cols. 8-9: stellar content extrapolated to $0.08\,M_{\odot}$.
\end{list}
\end{table*}

\begin{table*}
\caption[]{Stellar mass estimate for the clusters with PMS}
\label{tab8}
%\tiny
\renewcommand{\tabcolsep}{2.9mm}
\renewcommand{\arraystretch}{1.25}
\begin{tabular}{cccccccccccc}
\hline\hline
&&\multicolumn{4}{c}{MS}&&\multicolumn{2}{c}{PMS}&&\multicolumn{2}{c}{$MS+PMS$}\\
\cline{3-6}\cline{8-9}\cline{11-12}
Cluster&&$\Delta\,m_{MS}$&$\chi$&$N$&$M$   &&$N$&$M$ &&$N$&$M$\\
     && ($M_\odot$)    &      &(stars) &($M_\odot$)&&(stars)  &($M_\odot$)&&(stars) &($M_\odot$)     \\
(1)&&(2)&(3)&(4)&(5)&&(6)&(7)&&(8)&(9)\\
\hline
FSR\,807&&1.90-5.25&$-1.08\pm0.10$&$8\pm3$&$25\pm10$&&$21\pm6$&$45\pm14$&&$29\pm9$&$70\pm24$\\
FSR\,812&&6.75-13.0&$-0.34\pm0.15$&$9\pm3$&$83\pm30$&&$33\pm12$&$135\pm35$&&$42\pm15$&$218\pm65$\\
FSR\,826&&1.30-9.75&$1.07\pm0.12$&$46\pm9$&$178\pm31$&&$69\pm9$&$172\pm43$&&$115\pm18$&$350\pm74$\\
FSR\,904&&1.50-11.0&$0.26\pm0.16$&$61\pm8$&$188\pm42$&&$443\pm31$&$706\pm61$&&$504\pm39$&$894\pm103$\\
FSR\,955&&0.85-13.0&$0.22\pm0.09$&$4\pm2$&$20\pm7$&&$26\pm7$&$109\pm35$&&$30\pm9$&$129\pm42$\\
\hline
\end{tabular}
\begin{list}{Table Notes.}
\item Col. 2: MS mass range. Col. 3: MS mass function slope $\chi$, derived from the fit of $\phi(m)\propto{m}^{-(1+\chi)}$. Cols. 4-7: stellar content of the MS and PMS stars. Cols. 8-9: total (MS+PMS) stellar content.
\end{list}
\end{table*}

As a consequence of large-scale mass segregation, massive stars tend to be more concentrated in the core of evolved clusters, while low-mass stars are transferred to the outer regions \citep{Bonatto05}. As a consequence of mass segregation, $R_{RDP}$ tends to increase, while $R_{core}$ decreases.

\begin{figure}
\resizebox{\hsize}{!}{\includegraphics{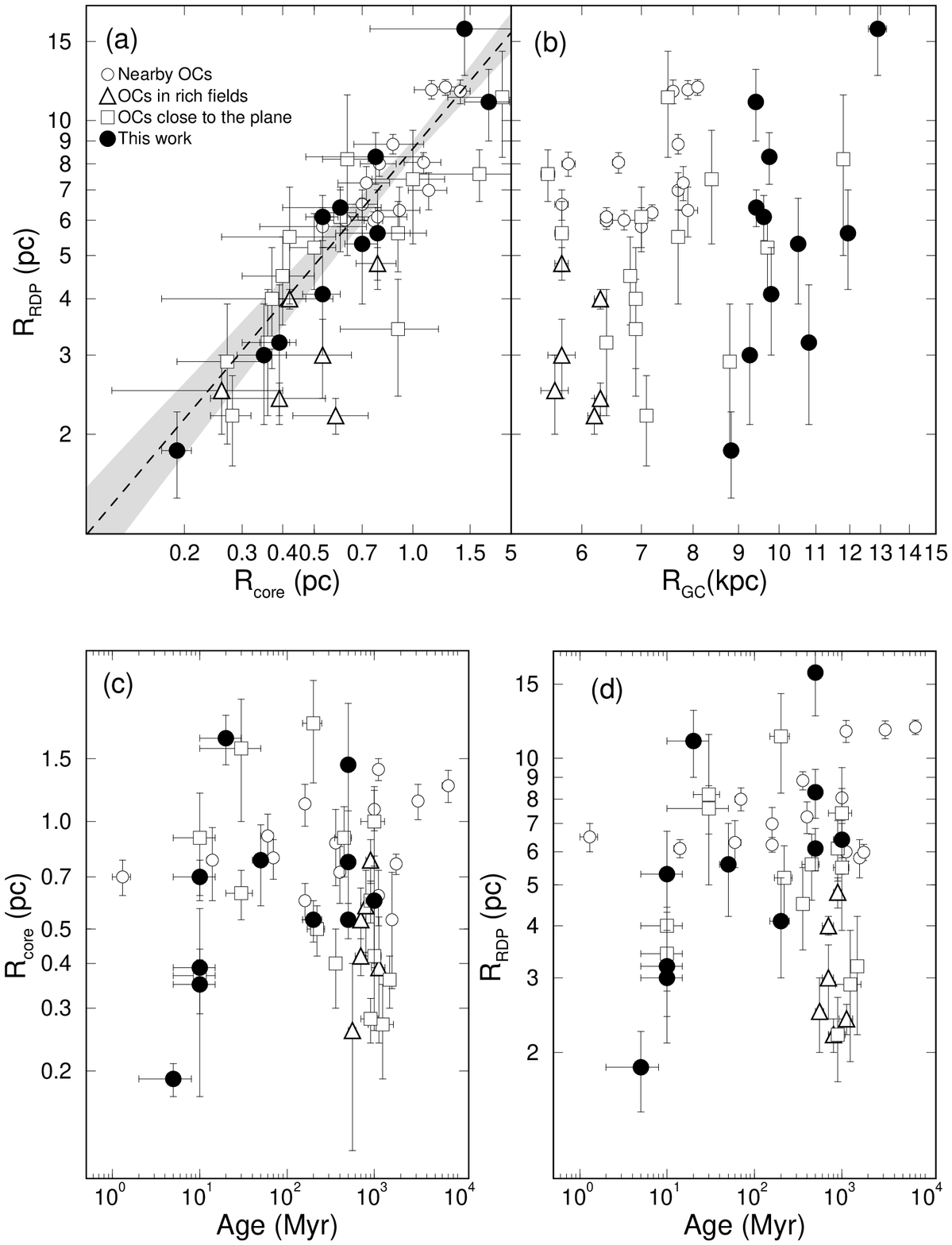}}
\caption[]{Relations involving structural parameters of OCs. Empty circles: nearby OCs, including two young ones. Triangles: OCs projected on dense fields towards the Galactic central regions. Square: OCs close to the Galactic plane. Filled circles: the present work OCs.}
\label{fig:13}
\end{figure}

In Fig. \ref{fig:13} we compare the structural parameters derived for the present OCs with those measured in different environments. As a reference sample, we use (\textit{i}) some bright nearby OCs \citep{Bonatto05}, including the two young OCs NGC 6611 \citep{Bonatto06a} and NGC 4755 \citep{Bonatto06b}; (\textit{ii}) OCs projected against the central parts of the Galaxy \citep{Bonatto07b}; (\textit{iii}) OCs projected close to the Galactic plane \citep{Camargo09}; and (\textit{iv}) the present sample.

In panel (\textit{a}) of Fig. \ref{fig:13}, core and cluster radii of the OCs in sample (\textit{i}) are almost linearly related by $R_{RDP}=(8.9\pm0.3)\times R_{core}^{(1.0\pm0.1)}$, which suggests that both kinds of radii undergo a similar scaling, in the sense that, on average, larger clusters tend to have larger cores. However, $\frac{1}{3}$ of the OCs in sample (\textit{ii}) do not follow that relation, which suggests that they are either intrinsically small or have been suffering important evaporation effects. The core and cluster radii in sample (\textit{iii}) and the OCs of this work (\textit{iv}) are consistent with the relation  at the $1\sigma$ level. A dependence of OC size on Galactocentric distance is shown in panel (\textit{b}), as previously suggested by \citet{Lynga82} and \citet{Tadross02}. The core and cluster radii of the OCs in this work (\textit{iv}) are consistent with the one in samples (\textit{ii}) and (\textit{iii}). Most OCs of our sample are located in the inner disk and close to the spiral arms, so they are consistent with sample (\textit{iii}) and with those located in crowded fields (\textit{ii}). To explain the increase of Globular Cluster radii with Galactocentric distance, \citet{vandenBergh91} suggest that part of the relation may be primordial, the higher molecular gas density in central Galactic regions may have created clusters with small radii. In addition, most clusters with small sizes are concentred near the Galactic plane, especially for $R_{GC}<9.5$ kpc and $z_{GC}<100$ pc \citep{Tadross02,Wielen71,Wielen75}. FSR 807, which appears exceedingly small for its Galactocentric distance, possibly presents this primordial effect, with $z_{GC}=0.0$ located close to Orion-Cygnus arm. In panels (\textit{c}) and (\textit{d}) we compare core and cluster radii with cluster age, respectively. This relationship is intimately related to cluster survival/dissociation rates. Both kinds of radii present a similar dependence on age, in which part of the clusters expand with time, while some seem to shrink. The bifurcation occurs at an age $\approx1$ Gyr. \citet{Mackey03} observed a similar effect for the core radii of LMC and SMC clusters and \citet{Mackey08} argue that this radius-age correlation has a dynamical origin. 
They attribute the trend to slow contraction in $R_{core}$ to a dynamical relaxation and/or core collapse. The expansion was attributed to mass loss from rapid stellar evolution in a cluster that is mass-segregated or otherwise centrally concentrated and to heating due to a significant population of retained stellar mass black holes that are scattered into the cluster halo or ejected from the cluster \citep{Mackey07,Merritt04}. We also note that the radii of the  clusters of our sample are related to the age similarly to the (\textit{leaky}) ones of \citet{Pfalzner09} and optical clusters of \citet{Maciejewski07}. 

\begin{figure}
\resizebox{\hsize}{!}{\includegraphics{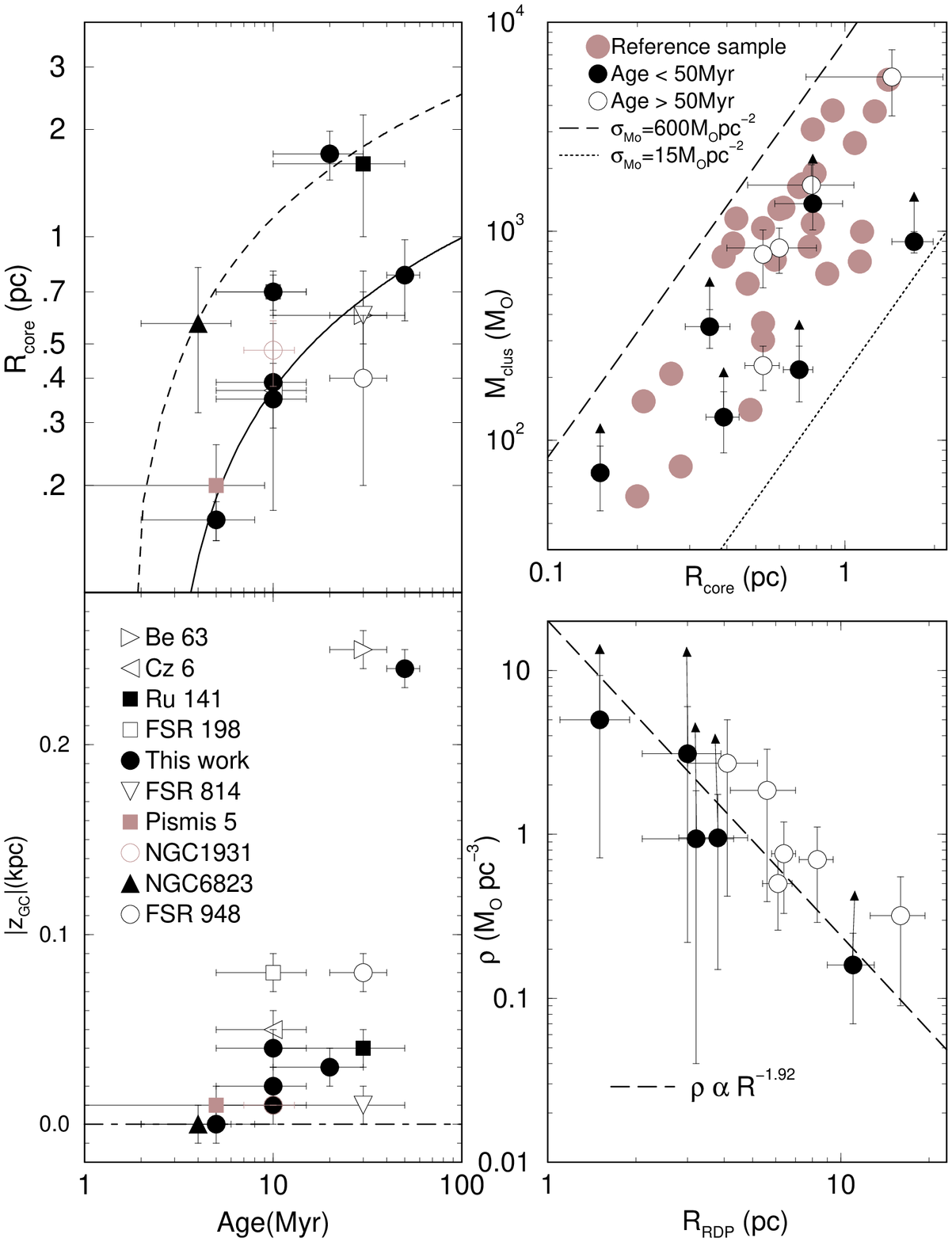}}
\caption[]{Top-left panel: age and $R_{core}$ relations. The dashed line is the logarithmic fit of \citet{Bastian08} 
($R_{core}(pc)=0.6\times{ln}(age[Myr])-0.25$) for M51 clusters. The solid line is our fit for Galactic clusters. 
Bottom-left panel: $|z_{GC}|$ for clusters of the top panel. Top right: core radius and cluster mass follow the 
relation $M_{clus}=13.8\sigma_{M0}R^{2}_{c}$, with varying values of $\sigma_{M0}$; reference sample OCs are shown
as grey circles. Filled circles show the OCs younger than 50\,Myr, while the older ones are shown as open circles. Bottom right: cluster density {\em vs.} radius.
Arrows indicate lower limits to the cluster mass and density.}
\label{fig:14}
\end{figure}

Most of the confirmed OCs of our sample are smaller than nearby OCs (\textit{i}) of similar age and Galactocentric distance (Fig.\ref{fig:13}). We point out that our sample occurs close to the Galactic plane and/or close to spiral arms (except NGC2234). Possibly, some of these OCs have interacted with GMCs. Molecular clouds more massive than $\sim10^{6}M_{\odot}$ are found in the solar neighbourhood \citep{Solomon87}. 
In Fig. \ref{fig:14} we show the relation between age and both $R_{core}$ and $|Z_{GC}|$ for the young OCs of our sample. We fit an empirical curve $R_c(pc)=0.27\times{ln}\,(age[Myr])-0.25$ to young OCs close to the Galactic plane and/or close to spiral arms. \citet{Bastian08} observed a similar relation for M51 clusters and fitted the empirical curve $R_c(pc)=0.6\times{ln}\,(age[Myr])-0.25$ that represents initially compact and mass-segregated star clusters ($R_{core}\approx0.25$ pc). In this context, N-body simulations by \citet{Vesperini09} show that in segregated clusters, early mass loss due to stellar evolution triggers a stronger expansion than for unsegregated clusters and that long-lived clusters initially with a high degree of mass segregation tend to have looser structure and reach core collapse later in their evolution than initially unsegregated clusters.

\begin{figure}
\resizebox{\hsize}{!}{\includegraphics{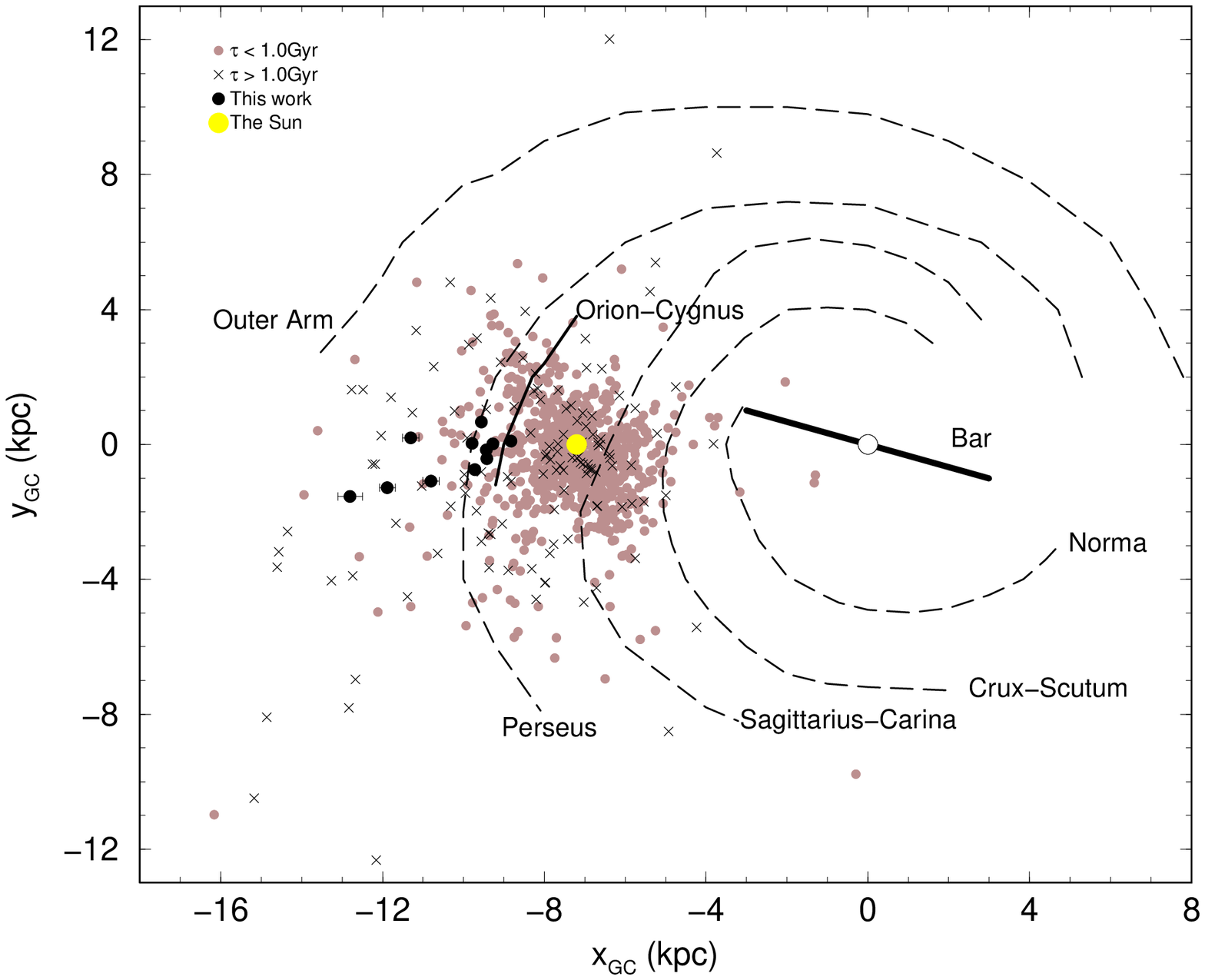}}
\caption[]{Spatial distribution of the present star clusters (filled circles) compared to the WEBDA OCs with ages younger than 1 Gyr (brown circles) and older than 1 Gyr (crosses). The schematic projection of the Galaxy is seen from the north pole, with 7.2 kpc as the Sun's distance to the Galactic centre.}
\label{fig:15}
\end{figure}

As discussed in \citet{Bonatto09c}, when the projected mass density of a star cluster 
follows a King-like profile (e.g. \citealt{StrucPar}), the cluster mass ($\rm M_{clus}$) 
can be expressed as a function of the core radius and the central surface mass density
($\sigma_{M0}$) according to $\rm M_{clus}\approx13.8\sigma_{M0}\,R^2_{C}$. Figure~\ref{fig:14}
(top-right panel) shows the distribution of our OCs in the plane core radius (Sect.~\ref{sec:4}) 
{\em vs} cluster mass (Sect.~\ref{Mass}). Clearly, our OCs (together with the reference sample)
distribute parallel to the above relation, since they are constrained within King-like distributions 
with $\rm15\la\sigma_{M0}\,(\ms\,pc^{-2})\la600$. (These bounds take the uncertainties into 
account.) This suggests a correlation between cluster mass and core radius, somewhat consistent 
with the mass-radius relation suggested by \citet{PZ10} for massive clusters younger than 
100\,Myr. 

Finally, with the cluster radius and mass estimates, we computed the cluster
mass density $\rm\rho(\ms\,pc^{-3})=\frac{3}{4\pi}M_{clus}\,R^{-3}_{RDP}$. We show the results
in the plane $R_{RDP}\, vs.\, \rho$ (Fig.~\ref{fig:14}, bottom-right panel). Despite the
error bars, the density decreases with cluster radius as $\rho\propto R^{-(1.92\pm0.36)}_{RDP}$, 
similarly to the sample of embedded clusters studied by \citet{Pfalzner09}. That work,
notes that the dependence on radius of the embedded clusters is significantly shallower 
than those observed in young clusters more massive than $10^3\,\ms$, $\rho\propto R^{-3}$ and
$R^{-4}$. As a caveat, we note that the mass density of the OCs in \citet{Pfalzner09} is more 
than 10 times higher than those of our OCs.

Figure \ref{fig:15} shows the spatial distribution in the Galactic plane and spiral arms \citep{Momany06} of the present OCs, compared to that of the OCs in the WEBDA database. We considered two age ranges, $<1$ Gyr and $>1$ Gyr, and computed the projections on the Galactic plane of the Galactic coordinates $(\ell,b)$. Old OCs are mainly found outside the Solar circle, and the inner Galaxy contains few OCs so far detected. The interesting point is whether inner Galaxy clusters cannot be observed because of strong absorption and crowding, or have been systematically dissolved by the different tidal effects combined \citep[][ and references therein]{Bonatto07a}. In this context, the more OCs are identified (with their astrophysical parameters derived) in the central parts, the more constraints can be established to settle this issue.

\section{Concluding remarks}
\label{sec:7}
We investigate the nature of 50 overdensities projected nearly towards the anti-centre, in the sector $160^\circ\,\leq\,\ell\,\leq 200^\circ$, with $|b|\,\leq\,20^\circ$ that were classified by \citet{Froebrich07} as probable OCs and labelled with quality flags 2 and 3.
The candidates are analysed by means of 2MASS colour-magnitude diagrams, stellar radial density profiles, and colour-colour diagrams for young objects. Field-star decontamination is applied to uncover the cluster's intrinsic CMD morphology, and CM filters are used to disentangle probable cluster members.

Out of the 50 overdensities, 16 (32\%) are confirmed as OCs. Nine (18\%) are new OCs (FSR 735, FSR 807, FSR 812, FSR 826, FSR 852, FSR 904, FSR 941, FSR 953, and FSR 955) and we derived astrophysical parameters. They are OCs or embedded clusters with age in the range 5 Myr to 1 Gyr, at distances from the Sun $1.28\lesssim{d}_{\odot}\lesssim5.78$ and Galactocentric distances $8.5\lesssim{R}_{GC}\lesssim12.9$. Other 7 (14\%) overdensities are previously catalogued OCs or embedded clusters (KKC1, FSR  795, Cz 22, FSR 828, FSR 856, Czernik 24, and NGC 2234). We also  derived parameters for Cz 22 and NGC2234. Five are classified as uncertain cases and require deeper photometry to establish their nature. The remaining FSR overdensities appear to be field fluctuations.

Most of the new OCs are located close to spiral arms and/or close to the Galactic plane and, probably because of this, the core radius appears to be smaller than the others at comparable Galactocentric distance and age. Also for this reason, most of them were undetected in the past.

\section*{Acknowledgements}
We thank an anonymous referee for significant comments and suggestions. This publication makes use of data products from the Two Micron All Sky Survey, which is a joint project of the University of Massachusetts and the Infrared Processing and Analysis Centre/California Institute of Technology, funded by the National Aeronautics and Space Administration and the National Science Foundation. This research has made use of the WEBDA database, operated at the Institute for Astronomy of the University of Vienna, as well as Digitised Sky Survey images from the Space Telescope Science Institute obtained using the extraction tool from CADC (Canada). We acknowledge support from CNPq and Capes (Brazil).

\end{document}